\documentclass{article}
\usepackage{graphicx} 

\usepackage[a4paper, margin=1in]{geometry}  
\usepackage{graphicx}  
\usepackage{titlesec}  
\usepackage{fancyhdr}  
\usepackage{hyperref}  
\usepackage{amsmath, amssymb}  
\usepackage{todonotes}  

\usepackage[edges]{forest}
\usepackage{xcolor}
\usepackage[skins]{tcolorbox}%
\usepackage{tcolorbox}
\usepackage{hyperref}
\newtcolorbox[auto counter, number within=section]{definitionbox}[2][]{%
colframe=black!50,
colback=black!5,
coltitle=white,
fonttitle=\bfseries,
title={#2},
sharp corners=south,
enhanced,
before upper={\noindent},
}

\newcommand{\ignore}[1]{}

\title{\textbf{AI Risk Atlas:\\
Taxonomy and Tooling for\\Navigating AI Risks and Resources}}

\author{Frank Bagehorn, Kristina Brimijoin, Elizabeth M. Daly, Jessica He,\\Michael Hind, Luis Garcés-Erice, Christopher Giblin, Ioana Giurgiu,\\Jacquelyn Martino, Rahul Nair, David Piorkowski, Ambrish Rawat, John Richards,\\Sean Rooney, Dhaval Salwala, Seshu Tirupathi, Peter Urbanetz,\\Kush R. Varshney, Inge Vejsbjerg, Mira L. Wolf-Bauwens}

\begin{document}

\maketitle  

\footnotetext{Authors are listed in alphabetical order by last name.}



\begin{abstract}
The rapid evolution of generative AI has expanded the breadth of risks associated with AI systems. While various taxonomies and frameworks exist to classify these risks, the lack of interoperability between them creates challenges for researchers, practitioners, and policymakers seeking to operationalise AI governance. To address this gap, we introduce the \textbf{AI Risk Atlas}, a structured taxonomy that consolidates AI risks from diverse sources and aligns them with governance frameworks. Additionally, we present the \textbf{Risk Atlas Nexus}, a collection of open-source tools designed to bridge the divide between risk definitions, benchmarks, datasets, and mitigation strategies. This knowledge-driven approach leverages ontologies and knowledge graphs to facilitate risk identification, prioritization, and mitigation. By integrating AI-assisted compliance workflows and automation strategies, our framework lowers the barrier to responsible AI adoption. We invite the broader research and open-source community to contribute to this evolving initiative, fostering cross-domain collaboration and ensuring AI governance keeps pace with technological advancements.
\end{abstract}

\vspace{1em}

\href{https://github.com/IBM/risk-atlas-nexus}{
  \includegraphics[height=1em]{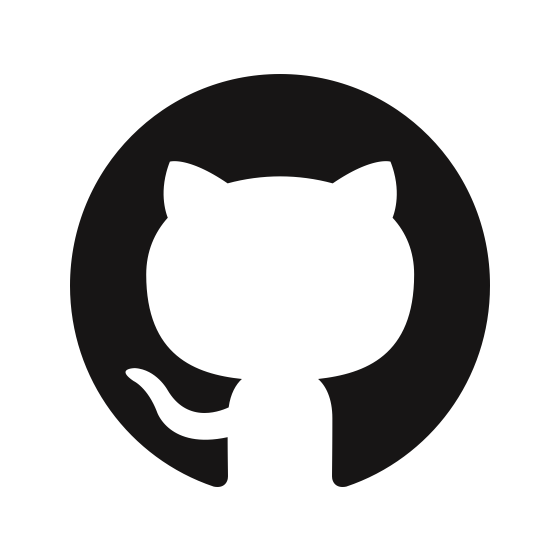}
  \hspace{0.5em}
  \texttt{https://github.com/IBM/risk-atlas-nexus}
}

\section{Introduction}
%
Identifying the risks of AI systems has attracted interest from research~\cite{weizenbaum1976computer, russell2016artificial}, industry~\cite{shahriari2017ieee}, and policy makers~\cite{carricco2018eu}.
These perspectives spawned many innovations to aid in the creation and operationalising of responsible AI system design \cite{aif360, arya2021ai, wexler2019if, bird2020fairlearn}. 
Generative AI and its rapidly evolving capabilities~\cite{bommasani2021opportunities}, increases the spectrum of risks, and the urgency to mitigate them. More recently, advent of agentic AI introduces new risks by giving AI automony and the ability to execute actions. As a result, there is a need to help the community identify and address these risks in parallel, while also creating a mechanism for collaboration.

There have been several efforts to catalogue risks associated with AI systems \cite{nist,owasp,mit-risk-repository,slattery2024ai,zeng2024ai}. However, connections and relationships to existing risk classification frameworks are missing. This lack of connectivity can present a challenge for practitioners who may want to adopt new risk taxonomies, but have already categorised their assets using existing definitions. 

%
Meanwhile, there are currently flourishing communities for sharing datasets~\cite{lhoest2021datasets} and benchmarks~\cite{eval-harness}. Although the focus of datasets effort is to evaluate the correctness of an AI model, typically partitioned by AI task or skill, many of these datasets can also be leveraged to assess various AI risks. Similarly, mapping the results of benchmarks to risk concerns is currently not a part of most benchmarks, but could be accomplished with better collaboration between the benchmark and risk communities~\cite{sokol2025benchmarkcardsstandardizeddocumentationlarge}.


%
To help manage risks, the process of putting a new AI system into production often includes multiple stakeholders such as business owners, risk and compliance officers, and ethics officers approving the AI system for a specific usage. Governance frameworks to manage this process typically include multiple manual steps, including curating information needed to assess risks (where will the system be used? who is the target user?) and reviewing outcomes to identify appropriate actions and governance strategies~\cite{piorkowski2025developingriskidentificationframework}.
Automation can play a vital role in easing the barrier for entry for developers and practitioners to make responsible AI an integral part of their process, however, this automation requires some structure of the underlying information.
For example, AI capabilities can help to create better semi-structured governance information, identify and prioritize risks according to the intended use case, recommend appropriate benchmarks and risk assessments and most importantly recommend mitigation strategies and recommended actions.  Our aim is to develop an AI Systems risk ontology that links
risk entities described using multiple different taxonomies with AI models,
evaluations, mitigations and other important entities. This ontology is manifested in a knowledge graph that associated tooling uses to integrate distinct risk frameworks, thereby providing the community with a way to align their assets with both new and existing risk definitions. 

This paper is organized as follows. Section~\ref{sec-risk-atlas} describes the AI Risk Atlas, which provides a taxonomy of AI Risks. Section~\ref{sec-tools} presents \textbf{Risk Atlas Nexus}, an open source tooling effort we have developed to enable inter- and cross-community collaboration. 
Section~\ref{sec-potential} discusses some of the future directions that can be facilitated by these tools.
Section~\ref{sec-conc} invites the broader community to expand on this initial seeding of tools by bringing different perspectives with different needs and ultimately lowering the barrier to AI governance.


\section{The AI Risk Landscape} \label{sec-risk-atlas}
This section describes the \textit{AI Risk Atlas} which aims to provide clarity for practitioners of the risks associated with Generative AI systems. 

The AI Risk Atlas is a taxonomy of AI risks collected from prior research, real-world examples, and from experts in the field. It defines risks posed by AI systems and explain potential consequences of those risks. Each risk is grouped into one of five categories based on where the risk originates. The categories are input risks, inference risks, output risks and non-technical risks. Within each category, risks are further grouped into risk dimensions such as accuracy, fairness, or explainability. These dimensions classify the individual risks into groups, and enable a user of the Atlas to focus on the dimensions relevant to them.

\begin{figure*}[!ht]
    \resizebox{0.98\textwidth}{!}{
    \begin{forest}
        forked edges, folder indent=0.5cm,
        where={level()<1}{}{folder, grow'=east},
        where={level()>0}{l sep+=0.55cm}{},
        for tree={
            fork sep=4mm,
            thick, edge=thick,
            font=\sffamily,
            if n children=0{if n=1{yshift=-5mm}{}, for parent={s sep=0mm}}{draw, minimum height=4ex, minimum width=4.0cm}
        }
        [AI Risk Atlas, calign=edge midpoint, 
            [Training data 
                [Transparency (2)]
                [Data laws (3)]
                [Privacy (3)]
                [Fairness (1)]
                [Intellectual property (2)]
                [Accuracy (2)]
                [Value alignment (2)]
                [Robustness (1)]
            ]
            [Inference 
                [Model-behavior manipulation (3)]
                [Prompt attacks (9)]
                [Privacy (3)]
                [Intellectual property (2)]
                [Accuracy (1)]
            ]
            [Output
                [Misuse (6)]
                [Value alignment (5)]
                [Intellectual property (2)]
                [Explainability (4)]
                [Robustness (1)]
                [Fairness (2)]
                [Privacy (1)]
            ]
            [Non-technical
                [Legal compliance (3)]
                [Governance (7)]
                [Societal impact (8)]
            ]
            [Agentic
                [Computational inefficiency (1)]
                [Explainability (1)]
                [Fairness (2)]
                [Governance (6)]
                [Privacy (2)]
                [Robustness (4)]
                [Societal impact (4)]
                [Value alignment (2)]         
            ]
        ]
    \end{forest}
}
    \caption{AI Risk Atlas Taxonomy. The taxonomy is divided into five categories, which are then further divided into dimensions. Next to each dimension in the parentheses is the number of risks identified for that dimension.}
    \label{fig:atlas-main}
\end{figure*}

\begin{figure*}
    \centering
    \includegraphics[width=\linewidth]{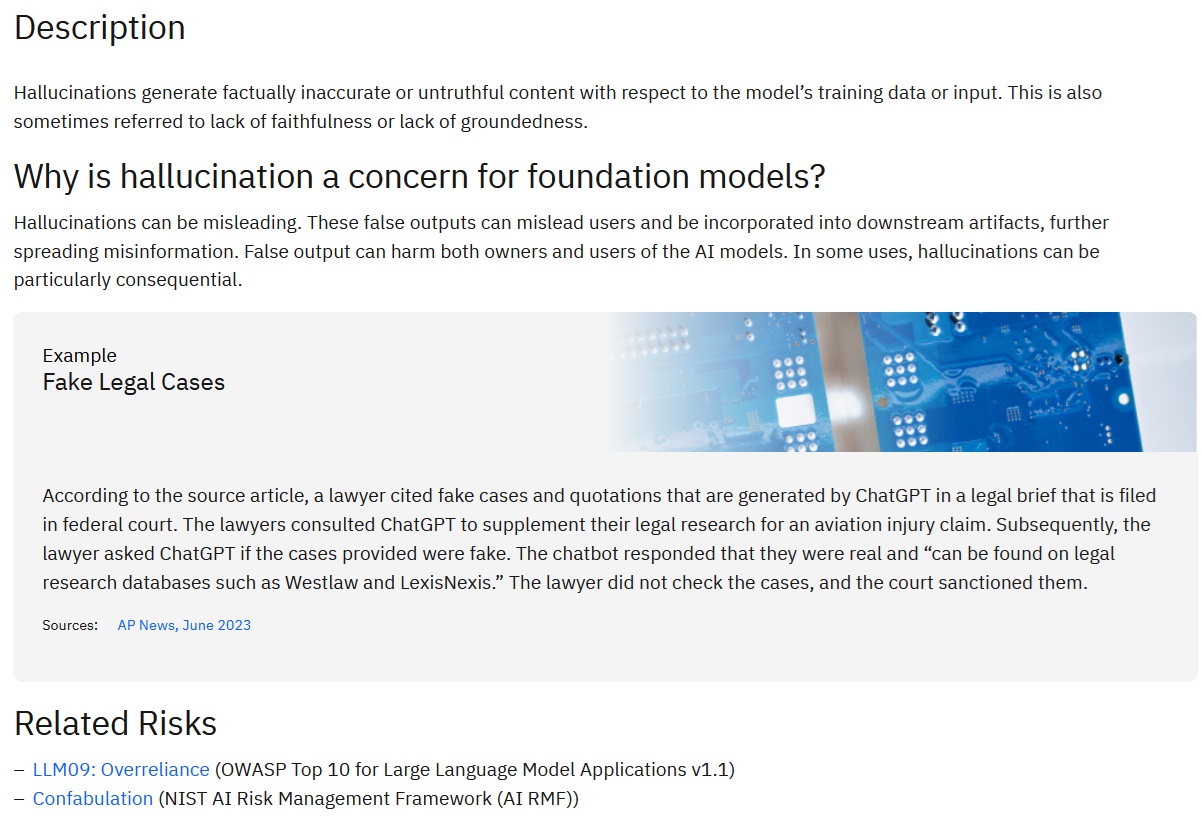}
    \caption{Screenshot of AI Risk Atlas Detail Page for Hallucination}
    \label{fig:atlas-detail-screenshot}
\end{figure*}

\ignore{
\begin{figure*}
\begin{definitionbox}{Hallucination}
Hallucinations generate factually inaccurate or untruthful content with respect to the model's training data or input. This is also sometimes referred to lack of faithfulness or lack of groundedness.\newline\newline
\textbf{Concern: }Hallucinations can be misleading. These false outputs can mislead users and be incorporated into downstream artifacts, further spreading misinformation. False output can harm both owners and users of the AI models. In some uses, hallucinations can be particularly consequential.\newline\newline
\textbf{Type: }output\newline
\textbf{Descriptor: }specific \newline\newline
\textbf{Implementation details: } \newline
ID: atlas-hallucination \newline
Tag: hallucination \newline
URI:  \href{https://www.ibm.com/docs/en/watsonx/saas?topic=SSYOK8/wsj/ai-risk-atlas/hallucination.html}{IBM AI Risk Atlas - Hallucination}\newline
\end{definitionbox}
\caption{AI Risk Atlas Entry Example }
\label{fig:atlas-detail}
\end{figure*}
}


Figure~\ref{fig:atlas-main} shows the overall AI Risk Atlas taxonomy where each category may include a subset of risk definitions. 
For example, Figure~\ref{fig:atlas-detail-screenshot} shows a screenshot of the Hallucination risk from the AI Risk Atlas, which is in the Robustness dimension in the Output category in Figure~\ref{fig:atlas-main}. The details contain a description of the risk along with a description of why the risk is a concern, a public example, when available, of the risk being manifested, and any related risks in other popular taxonomies.

The creation of the AI Risk Atlas was motivated by the changing risk landscape due to the emergence and rapid success of generative AI. Before generative AI became ubiquitous, IBM Research had developed the techniques to evaluate and mitigate some risks such as fairness, explainability, adversarial robustness, privacy, and uncertainty for traditional (non-generative) models. This work evolved into open-source toolkits designed to measure, and in some cases, mitigate those risks~\cite{aif360,aix360,art360,aip360,uq360}. However, these toolkits did not initially account for the risks specific to generative models. The large amount of data used to train generative AI, the increasing complexity of the models, and the non-determinism prevalent were some of the factors that led to the identification of new risks for generative models. Prompt-based attacks, poorly curated training data, and hallucinations emerged as early risks identified by researchers and concerned citizens alike. The identification of risks had begun to outpace mechanisms for measuring and understanding the risks. 


To address this gap and to provide a foundation for understanding the risks of both traditional and generative AI models, 
IBM's AI Ethics Board created white papers~\cite{ethics-board-pov, ethics-board-pov-agents} that provided the foundation for the AI Risk Atlas.
The goal for the Atlas was multi-faceted. First, we wanted to have a single source of information for currently known AI risks. This would enable a shared vocabulary when discussing AI risks. Second, we wanted to identify opportunities for measuring and mitigating the newly identified risks. This would help identify research opportunities for underrepresented risks. Third, we wanted a resource to develop usage-based governance. Specifically, we wanted to address the question of what risks are relevant to particular use cases.

The AI Risk Atlas has been used as a conversation starting point with enterprises who are considering deploying AI. It helps these organizations to be aware of the possible risks they need to govern. It provides a palette or vocabulary of risks that enterprises can consider: they can decide the risk is relevant to their use case and develop a plan to mitigate the risk either with tooling, human oversight, or both.
For those risks that aren't applicable for a use case, the organization can document this risk to help demonstrate their risk governance framework. However, the AI Risk Atlas can be used further than just a conversation starter. It can provide the underlying vocabulary for the complete management of the risk from development to deployment to monitoring \cite{watsonx-gov-dec-2024,mra-vision, daly2024usage}.


\section{Tools for Practitioners} \label{sec-tools}
The IBM AI Risk Atlas has been used many enterprise customers to help them reason about the risks in their AI systems. In order to enable efforts to leverage these risks to  operationalise governance and risk mitigation frameworks we created \textit{Risk Atlas Nexus.}

The \textit{Risk Atlas Nexus} is a collection of tooling to help bring together disparate resources related to governance of foundation models. We aim to support a community-driven approach to curating and cataloguing resources such as datasets, benchmarks and mitigations. Our goal is to turn abstract risk definitions into actionable workflows that streamline AI governance processes. By connecting fragmented resources, Risk Atlas Nexus seeks to fill a critical gap in AI governance, enabling stakeholders to build more robust, transparent, and accountable systems. The Risk Atlas Nexus is a step towards enabling the following.

\textbf{Navigating disparate risk taxonomies:} IBM AI Risk Atlas is one amongst a number of existing risk taxonomies, for example; the OWASP Top 10 for LLMs and Generative AI Apps \cite{owasp}, the NIST AI Risk Management Framework \cite{nist}, the MIT AI Risk Repository \cite{mit-risk-repository, slattery2024ai}, the AIR taxonomy 2024 \cite{zeng2024ai}.

To provide a way through this labyrinth of taxonomies, we have constructed an AI risk ontology that allows both the creation of a knowledge graph containing those different taxonomies and the ability to map between them, leveraging the AIRO \cite{golpayegani2022airo}.The ontology has been modeled using LinkML \cite{moxon2021linked}, which allows the generation of different data representations (e.g. RDF, OWL) in a simple way. The risk taxonomies have been stored as LinkML data instance YAML files. To express some semantically meaningful mapping between risks from different taxonomies, we have used the Simple Standard for Sharing Ontological Mappings (SSSOM) \cite{sssom}. Therefore those mappings are maintained in SSOM TSV files and are converted to LinkML data YAML using Python helper scripts.

Sample notebooks demonstrate how to load the LinkML data and user data and how to get details about specific risks and their relations to risks in other taxonomies.

\textbf{Question Answering:} Compliance questionnaires are usually required prior to deploying an AI model into production. These enable a thorough understanding of the specific use case and associated risk exposures~\cite{watsonx-gov-dec-2024,lee2023qb4airaquestionbankai}. The Risk Atlas Nexus supports the development and curation of questionnaires to a desired taxonomy. Additionally, the content can support Large Language Models (LLMs) to assist users in responding to time-consuming compliance questionnaires, thereby reducing manual effort and minimizing errors~\cite{daly2024usage}. Similarly, other aspects like risk identification, guardrail implementation, and identifying security vulnerabilities for specific use cases can be largely automated with human feedback and sign-off provided only when necessary.

\textbf{Use Case to Risk Prioritisation:} Risk Atlas Nexus supports risk classificated according to the EU AI Act \cite{herdel2024exploregen}. Additionally, to help prioritise which of the many risks are most related to their use case we leverage LLM-as-a-judge capabilities to identify which risks to consider. This information can be used to look for appropriate research papers, benchmarks and metrics. In a similar manner Risk Atlas Nexus can be used to tag disparate resources by passing in text such as a paper abstract or a dataset description as the risks for the basis for an LLM-as-a-judge definition \cite{ashktorab2024aligning,desmond2025evalassist}.

\textbf{From Risks to Mitigating Actions:} The knowledge graph supports mapping between risks to two types of mitigation strategies: detectors and recommended actions. Detectors such as Granite Guardian~\cite{padhi2024graniteguardian} dimensions could be run in tandem with an AI system to better protect against certain risks such as social bias and prompt injection attacks. We have also mined recommended actions as part of the NIST AI Risk Management  Framework~\cite{nist} to be able to recommend more process driven mitigation strategies. Additionally, Risk Atlas Nexus supports linkages between benchmarks and evaluations associated with uncovering risks. 

\textbf{Bring Your Own Risks, Relationships and Questionnaires:} Risk Atlas Nexus tooling supports several well known risk taxonomy frameworks, however some organisations may wish to define their own custom concerns and definitions.  Risk Atlas Nexus allows users to define custom questionnaire templates as well as taxonomies, risks, mappings, and mitigation actions which should conform to the  \href{ https://github.com/IBM/risk-atlas-nexus/blob/42a42aebf87cdf18232105ab57ffef69331e322d/docs/ontology/index.md }{ontology schema}. We encourage users to contribute their taxonomy definitions and mappings back to the project for others to use through the open-source project.  

\textbf{GAF-Guard:} Given the evolving capabilities, agentic frameworks\cite{lang-graph,crewai,autogen,bee-ai-framework} have the potential to create real-time governance pipelines, from identifying relevant risks and benchmarks to identifying mitigating actions to online monitoring capabilities. GAF-Guard is an agentic framework that leverages Risk Atlas Nexus functionalities like auto assisted question answering, use-case to risk identification and prioritization. GAF-Guard agents perform these tasks and consequently automate the process of the governance life-cycle. GAF-Guard agents initially identify the risks associated with a given use-case and then generate risk monitors and detectors for near real-time monitoring of deployed LLMs for the given use-case.

\section{Potential for the future} \label{sec-potential}
There is immense potential of automation of various aspects of compliance and risk management processes. While human oversight and manual verification are still essential requirements for compliance, auditing and regulatory purposes, automation can help to bring an efficient execution of intermediate stages in the compliance workflow. AutoML strategies have been employed to automate model training pipelines excelling at tasks such as feature selection,  hyper parameter optimization, model generation and evaluation \cite{he2021automl,wang2020autoai}. In a similar manner AI governance pipelines could be employed to detect and mitigate risks. By starting with identifying the most relevant risks, running the most relevant benchmarks and then assessing the impact of employing real-time mitigation strategies, concrete recommendations can be made to improve safety of AI solutions.

In the context of complex systems LLMs are increasingly being used as part of the validation process, from software testing and assisting in tasks such as test case preparation \cite{wang2024software} to assessing the output of an LLM \cite{van2024field,shankar2024validates, ashktorab2024aligning}. With the increasing capabilities of LLMs their applications have gone beyond single function tasks to being used to address complex problems acting as autonomous-agents \cite{wang2024survey}. ToolLLaMA learns how to call appropriate API based tools \cite{schick2023toolformer} meaning LLMs can orchestrate tasks that leverage existing functionality embedded in other tooling. The research community has begun to use agent flows to design, plan and execute scientific experiments  \cite{boiko2023emergent} and even write papers \cite{lu2024ai}.

\section{Conclusion and call to action} \label{sec-conc}

Curating the AI Risk Atlas is just the first step in providing a reference framework for researchers and practitioners navigating the rapidly evolving AI landscape. By positioning our risk taxonomy in relation to existing definitions and taxonomies, we aim to encourage the community to map new risk definitions, datasets, benchmarks, research papers, and crucially mitigation and detection strategies into a structured framework. This approach will enhance accessibility and facilitate the operationalisation of AI governance processes.

The initial tools released as part of the Risk Atlas Nexus toolkit represent only the beginning of what is possible. We are committed to ongoing development, enabling the developer community to contribute and expand this initiative. By fostering a community-driven approach, we can lower the barrier to entry for all. We invite the open-source community to enrich the knowledge graph by linking benchmarks, datasets, and research papers to identified risks. Additionally, contributors can request new functionality through GitHub enhancement requests or develop and integrate their own algorithms.

\bibliographystyle{plain}
\bibliography{refs}

\appendix
\section{AI Risk Atlas Definitions}
The below is a catalog of potential risks when working with generative AI, foundation models, and machine learning models.

\begin{definitionbox}{Evasion attack}
Evasion attacks attempt to make a model output incorrect results by slightly perturbing the input data sent to the trained model.\newline\newline
\textbf{Concern: }Evasion attacks alter model behavior, usually to benefit the attacker.\newline\newline
\textbf{Type: }inference\newline
\textbf{Descriptor: }amplified by generative AI \newline\newline
\textbf{Implementation details: } \newline
ID: atlas-evasion-attack \newline
Tag: evasion-attack \newline
URI:  \href{https://www.ibm.com/docs/en/watsonx/saas?topic=SSYOK8/wsj/ai-risk-atlas/evasion-attack.html}{IBM AI Risk Atlas - Evasion attack}\newline
\end{definitionbox}
\begin{definitionbox}{Impact on the environment}
AI, and large generative models in particular, might produce increased carbon emissions and increase water usage for their training and operation.\newline\newline
\textbf{Concern: }Training and operating large AI models, building data centers, and manufacturing specialized hardware for AI can consume large amounts of water and energy, which contributes to carbon emissions. Additionally, water resources that are used for cooling AI data center servers can no longer be allocated for other necessary uses. If not managed, these could exacerbate climate change. \newline\newline
\textbf{Type: }non-technical\newline
\textbf{Descriptor: }amplified by generative AI \newline\newline
\textbf{Implementation details: } \newline
ID: atlas-impact-on-the-environment \newline
Tag: impact-on-the-environment \newline
URI:  \href{https://www.ibm.com/docs/en/watsonx/saas?topic=SSYOK8/wsj/ai-risk-atlas/impact-on-the-environment.html}{IBM AI Risk Atlas - Impact on the environment}\newline
\end{definitionbox}
\begin{definitionbox}{Incorrect risk testing}
A metric selected to measure or track a risk is incorrectly selected, incompletely measuring the risk, or measuring the wrong risk for the given context.\newline\newline
\textbf{Concern: }If the metrics do not measure the risk as intended, then the understanding of that risk will be incorrect and mitigations might not be applied. If the model's output is consequential, this might result in societal, reputational, or financial harm.\newline\newline
\textbf{Type: }non-technical\newline
\textbf{Descriptor: }amplified by generative AI \newline\newline
\textbf{Implementation details: } \newline
ID: atlas-incorrect-risk-testing \newline
Tag: incorrect-risk-testing \newline
URI:  \href{https://www.ibm.com/docs/en/watsonx/saas?topic=SSYOK8/wsj/ai-risk-atlas/incorrect-risk-testing.html}{IBM AI Risk Atlas - Incorrect risk testing}\newline
\end{definitionbox}
\begin{definitionbox}{Over- or under-reliance}
In AI-assisted decision-making tasks, reliance measures how much a person trusts (and potentially acts on) a model's output. Over-reliance occurs when a person puts too much trust in a model, accepting a model's output when the model's output is likely incorrect. Under-reliance is the opposite, where the person doesn't trust the model but should.\newline\newline
\textbf{Concern: }In tasks where humans make choices based on AI-based suggestions, over/under reliance can lead to poor decision making because of the misplaced trust in the AI system, with negative consequences that increase with the importance of the decision.\newline\newline
\textbf{Type: }output\newline
\textbf{Descriptor: }amplified by generative AI \newline\newline
\textbf{Implementation details: } \newline
ID: atlas-over-or-under-reliance \newline
Tag: over-or-under-reliance \newline
URI:  \href{https://www.ibm.com/docs/en/watsonx/saas?topic=SSYOK8/wsj/ai-risk-atlas/over-or-under-reliance.html}{IBM AI Risk Atlas - Over- or under-reliance}\newline
\end{definitionbox}
\begin{definitionbox}{Membership inference attack}
A membership inference attack repeatedly queries a model to determine if a given input was part of the model's training. More specifically, given a trained model and a data sample, an attacker appropriately samples the input space, observing outputs to deduce whether that sample was part of the model's training.\newline\newline
\textbf{Concern: }Identifying whether a data sample was used for training data can reveal what data was used to train a model, possibly giving competitors insight into how a model was trained and the opportunity to replicate the model or tamper with it. Models that include publicly-available data are at higher risk of such attacks.\newline\newline
\textbf{Type: }inference\newline
\textbf{Descriptor: }amplified by generative AI \newline\newline
\textbf{Implementation details: } \newline
ID: atlas-membership-inference-attack \newline
Tag: membership-inference-attack \newline
URI:  \href{https://www.ibm.com/docs/en/watsonx/saas?topic=SSYOK8/wsj/ai-risk-atlas/membership-inference-attack.html}{IBM AI Risk Atlas - Membership inference attack}\newline
\end{definitionbox}
\begin{definitionbox}{Confidential data in prompt}
Confidential information might be included as a part of the prompt that is sent to the model.\newline\newline
\textbf{Concern: }If not properly developed to secure confidential data, the model might reveal confidential information or IP in the generated output. Additionally, end users' confidential information might be unintentionally collected and stored.\newline\newline
\textbf{Type: }inference\newline
\textbf{Descriptor: }specific to generative AI \newline\newline
\textbf{Implementation details: } \newline
ID: atlas-confidential-data-in-prompt \newline
Tag: confidential-data-in-prompt \newline
URI:  \href{https://www.ibm.com/docs/en/watsonx/saas?topic=SSYOK8/wsj/ai-risk-atlas/confidential-data-in-prompt.html}{IBM AI Risk Atlas - Confidential data in prompt}\newline
\end{definitionbox}
\begin{definitionbox}{Prompt leaking}
A prompt leak attack attempts to extract a model's system prompt (also known as the system message).\newline\newline
\textbf{Concern: }A successful prompt leaking attack copies the system prompt used in the model. Depending on the content of that prompt, the attacker might gain access to valuable information, such as sensitive personal information or intellectual property, and might be able to replicate some of the functionality of the model.\newline\newline
\textbf{Type: }inference\newline
\textbf{Descriptor: }specific to generative AI \newline\newline
\textbf{Implementation details: } \newline
ID: atlas-prompt-leaking \newline
Tag: prompt-leaking \newline
URI:  \href{https://www.ibm.com/docs/en/watsonx/saas?topic=SSYOK8/wsj/ai-risk-atlas/prompt-leaking.html}{IBM AI Risk Atlas - Prompt leaking}\newline
\end{definitionbox}
\begin{definitionbox}{Data privacy rights alignment}
Existing laws could include providing data subject rights such as opt-out, right to access, and right to be forgotten.\newline\newline
\textbf{Concern: }Improper usage or a request for data removal could force organizations to retrain the model, which is expensive.\newline\newline
\textbf{Type: }training-data\newline
\textbf{Descriptor: }amplified by generative AI \newline\newline
\textbf{Implementation details: } \newline
ID: atlas-data-privacy-rights \newline
Tag: data-privacy-rights \newline
URI:  \href{https://www.ibm.com/docs/en/watsonx/saas?topic=SSYOK8/wsj/ai-risk-atlas/data-privacy-rights.html}{IBM AI Risk Atlas - Data privacy rights alignment}\newline
\end{definitionbox}
\begin{definitionbox}{Discriminatory actions}
AI agents can take actions where one group of humans is unfairly advantaged over another due to the decisions of the model. This may be caused by AI agents' biased actions that impact the world, in the resources consulted, and in the resource selection process. For example, an AI agent can generate code that can be biased.\newline\newline
\textbf{Concern: }Discriminatory actions can cause harm to people. Discriminatory actions taken by an AI agent could perpetuate bias to systems outside the AI agent owner's control,  impact people, or lead to unintended consequences.\newline\newline
\textbf{Type: }agentic\newline
\textbf{Descriptor: }amplified by agentic AI \newline\newline
\textbf{Implementation details: } \newline
ID: atlas-discriminatory-actions \newline
Tag: discriminatory-actions \newline
URI:  \href{https://www.ibm.com/docs/en/watsonx/saas?topic=SSYOK8/wsj/ai-risk-atlas/discriminatory-actions.html}{IBM AI Risk Atlas - Discriminatory actions}\newline
\end{definitionbox}
\begin{definitionbox}{IP information in prompt}
Copyrighted information or other intellectual property might be included as a part of the prompt that is sent to the model.\newline\newline
\textbf{Concern: }Inclusion of such data might result in it being disclosed in the model output. In addition to accidental disclosure, prompt data might be used for other purposes like model evaluation and retraining, and might appear in their output if not properly removed.\newline\newline
\textbf{Type: }inference\newline
\textbf{Descriptor: }specific to generative AI \newline\newline
\textbf{Implementation details: } \newline
ID: atlas-ip-information-in-prompt \newline
Tag: ip-information-in-prompt \newline
URI:  \href{https://www.ibm.com/docs/en/watsonx/saas?topic=SSYOK8/wsj/ai-risk-atlas/ip-information-in-prompt.html}{IBM AI Risk Atlas - IP information in prompt}\newline
\end{definitionbox}
\begin{definitionbox}{Legal accountability}
Determining who is responsible for an AI model is challenging without good documentation and governance processes.\newline\newline
\textbf{Concern: }If ownership for development of the model is uncertain, regulators and others might have concerns about the model. It would not be clear who would be liable and responsible for the problems with it or can answer questions about it. Users of models without clear ownership might find challenges with compliance with future AI regulation.\newline\newline
\textbf{Type: }non-technical\newline
\textbf{Descriptor: }amplified by generative AI \newline\newline
\textbf{Implementation details: } \newline
ID: atlas-legal-accountability \newline
Tag: legal-accountability \newline
URI:  \href{https://www.ibm.com/docs/en/watsonx/saas?topic=SSYOK8/wsj/ai-risk-atlas/legal-accountability.html}{IBM AI Risk Atlas - Legal accountability}\newline
\end{definitionbox}
\begin{definitionbox}{Hallucination}
Hallucinations generate factually inaccurate or untruthful content with respect to the model's training data or input. This is also sometimes referred to lack of faithfulness or lack of groundedness.\newline\newline
\textbf{Concern: }Hallucinations can be misleading. These false outputs can mislead users and be incorporated into downstream artifacts, further spreading misinformation. False output can harm both owners and users of the AI models. In some uses, hallucinations can be particularly consequential.\newline\newline
\textbf{Type: }output\newline
\textbf{Descriptor: }specific to generative AI \newline\newline
\textbf{Implementation details: } \newline
ID: atlas-hallucination \newline
Tag: hallucination \newline
URI:  \href{https://www.ibm.com/docs/en/watsonx/saas?topic=SSYOK8/wsj/ai-risk-atlas/hallucination.html}{IBM AI Risk Atlas - Hallucination}\newline
\end{definitionbox}
\begin{definitionbox}{Social hacking attack}
Manipulative prompts that use social engineering techniques, such as role-playing or hypothetical scenarios, to persuade the model into generating harmful content.\newline\newline
\textbf{Concern: }Social hacking attacks can be used to alter model behavior and benefit the attacker. The content it generates may cause harms for the user or others.\newline\newline
\textbf{Type: }inference\newline
\textbf{Descriptor: }specific to generative AI \newline\newline
\textbf{Implementation details: } \newline
ID: atlas-social-hacking-attack \newline
Tag: social-hacking-attack \newline
URI:  \href{https://www.ibm.com/docs/en/watsonx/saas?topic=SSYOK8/wsj/ai-risk-atlas/social-hacking-attack.html}{IBM AI Risk Atlas - Social hacking attack}\newline
\end{definitionbox}
\begin{definitionbox}{Harmful output}
A model might generate language that leads to physical harm. The language might include overtly violent, covertly dangerous, or otherwise indirectly unsafe statements.\newline\newline
\textbf{Concern: }A model generating harmful output can cause immediate physical harm or create prejudices that might lead to future harm.\newline\newline
\textbf{Type: }output\newline
\textbf{Descriptor: }specific to generative AI \newline\newline
\textbf{Implementation details: } \newline
ID: atlas-harmful-output \newline
Tag: harmful-output \newline
URI:  \href{https://www.ibm.com/docs/en/watsonx/saas?topic=SSYOK8/wsj/ai-risk-atlas/harmful-output.html}{IBM AI Risk Atlas - Harmful output}\newline
\end{definitionbox}
\begin{definitionbox}{Indirect instructions attack}
Prompts, questions, or requests designed to elicit undesirable responses from the application. Unlike direct instructions attacks, the model is instructed to use instructions that are embedded in external data like a website.\newline\newline
\textbf{Concern: }Indirect instructions attacks can be used to alter model behavior and benefit the attacker. The content it generates may cause harms for the user or others.\newline\newline
\textbf{Type: }inference\newline
\textbf{Descriptor: }specific to generative AI \newline\newline
\textbf{Implementation details: } \newline
ID: atlas-indirect-instructions-attack \newline
Tag: indirect-instructions-attack \newline
URI:  \href{https://www.ibm.com/docs/en/watsonx/saas?topic=SSYOK8/wsj/ai-risk-atlas/indirect-instructions-attack.html}{IBM AI Risk Atlas - Indirect instructions attack}\newline
\end{definitionbox}
\begin{definitionbox}{Mitigation and maintenance}
The large number of components and dependencies that agent systems have complicates keeping them up to date and correcting problems.\newline\newline
\textbf{Concern: }AI agents may interact with other systems, tools, or other agents. Tracing the root cause of failure becomes more difficult and more costly as agent capabilities and complexities increase.\newline\newline
\textbf{Type: }agentic\newline
\textbf{Descriptor: }amplified by agentic AI \newline\newline
\textbf{Implementation details: } \newline
ID: atlas-mitigation-maintenance \newline
Tag: mitigation-maintenance \newline
URI:  \href{https://www.ibm.com/docs/en/watsonx/saas?topic=SSYOK8/wsj/ai-risk-atlas/mitigation-maintenance.html}{IBM AI Risk Atlas - Mitigation and maintenance}\newline
\end{definitionbox}
\begin{definitionbox}{AI agent compliance}
Determining AI agents' compliance is complex and there might not be enough information to assess whether the agentic AI system is compliant with applicable legal requirements.\newline\newline
\textbf{Concern: }AI agents may interact with other systems, tools, or other agents. AI agents can also find solutions to accomplish a task or a goal in a variety of ways and there could be uncertainty around the way an AI agent would choose each time to perform the task. Assessing compliance can become more difficult as agent capabilities increase.\newline\newline
\textbf{Type: }agentic\newline
\textbf{Descriptor: }amplified by agentic AI \newline\newline
\textbf{Implementation details: } \newline
ID: atlas-ai-agent-compliance \newline
Tag: ai-agent-compliance \newline
URI:  \href{https://www.ibm.com/docs/en/watsonx/saas?topic=SSYOK8/wsj/ai-risk-atlas/ai-agent-compliance.html}{IBM AI Risk Atlas - AI agent compliance}\newline
\end{definitionbox}
\begin{definitionbox}{Function calling hallucination}
AI agents might make mistakes when generating function calls (calls to tools to execute actions). Those function calls might result in incorrect, unnecessary or harmful actions. Examples: Generating wrong functions or wrong parameters for the functions.\newline\newline
\textbf{Concern: }Hallucinations when generating function calls might result in wrong or redundant actions being performed. Depending on the actions taken, AI agents can cause harms to owners and users of the AI agents.\newline\newline
\textbf{Type: }agentic\newline
\textbf{Descriptor: }specific to agentic AI \newline\newline
\textbf{Implementation details: } \newline
ID: atlas-function-calling-hallucination \newline
Tag: function-calling-hallucination \newline
URI:  \href{https://www.ibm.com/docs/en/watsonx/saas?topic=SSYOK8/wsj/ai-risk-atlas/function-calling-hallucination.html}{IBM AI Risk Atlas - Function calling hallucination}\newline
\end{definitionbox}
\begin{definitionbox}{Confidential information in data}
Confidential information might be included as part of the data that is used to train or tune the model.\newline\newline
\textbf{Concern: }If confidential data is not properly protected, there could be an unwanted disclosure of confidential information. The model might expose confidential information in the generated output or to unauthorized users.\newline\newline
\textbf{Type: }training-data\newline
\textbf{Descriptor: }amplified by generative AI \newline\newline
\textbf{Implementation details: } \newline
ID: atlas-confidential-information-in-data \newline
Tag: confidential-information-in-data \newline
URI:  \href{https://www.ibm.com/docs/en/watsonx/saas?topic=SSYOK8/wsj/ai-risk-atlas/confidential-information-in-data.html}{IBM AI Risk Atlas - Confidential information in data}\newline
\end{definitionbox}
\begin{definitionbox}{Lack of model transparency}
Lack of model transparency is due to insufficient documentation of the model design, development, and evaluation process and the absence of insights into the inner workings of the model.\newline\newline
\textbf{Concern: }Transparency is important for legal compliance, AI ethics, and guiding appropriate use of models. Missing information might make it more difficult to evaluate risks,  change the model, or reuse it.  Knowledge about who built a model can also be an important factor in deciding whether to trust it. Additionally, transparency regarding how the model's risks were determined, evaluated, and mitigated also play a role in determining model risks, identifying model suitability, and governing model usage.\newline\newline
\textbf{Type: }non-technical\newline
\textbf{Descriptor: }traditional risk of AI \newline\newline
\textbf{Implementation details: } \newline
ID: atlas-lack-of-model-transparency \newline
Tag: lack-of-model-transparency \newline
URI:  \href{https://www.ibm.com/docs/en/watsonx/saas?topic=SSYOK8/wsj/ai-risk-atlas/lack-of-model-transparency.html}{IBM AI Risk Atlas - Lack of model transparency}\newline
\end{definitionbox}
\begin{definitionbox}{Exploit trust mismatch}
Attackers might initiate injection attacks to bypass the trust boundary, which is a distinct point or conceptual line where the level of trust in a system, application or network changes. Background execution in multi-agent environments increases the risk of covert channels if input/output validation is weak.\newline\newline
\textbf{Concern: }This could lead to mismatched (expected vs. realized) trust boundaries and could result in unintended tool use, excessive agency, and privilege escalation.\newline\newline
\textbf{Type: }agentic\newline
\textbf{Descriptor: }amplified by agentic AI \newline\newline
\textbf{Implementation details: } \newline
ID: atlas-exploit-trust-mismatch \newline
Tag: exploit-trust-mismatch \newline
URI:  \href{https://www.ibm.com/docs/en/watsonx/saas?topic=SSYOK8/wsj/ai-risk-atlas/exploit-trust-mismatch.html}{IBM AI Risk Atlas - Exploit trust mismatch}\newline
\end{definitionbox}
\begin{definitionbox}{Unrepresentative data}
Unrepresentative data occurs when the training or fine-tuning data is not sufficiently representative of the underlying population or does not measure the phenomenon of interest.\newline\newline
\textbf{Concern: }If the data is not representative, then the model will not work as intended.\newline\newline
\textbf{Type: }training-data\newline
\textbf{Descriptor: }traditional risk of AI \newline\newline
\textbf{Implementation details: } \newline
ID: atlas-unrepresentative-data \newline
Tag: unrepresentative-data \newline
URI:  \href{https://www.ibm.com/docs/en/watsonx/saas?topic=SSYOK8/wsj/ai-risk-atlas/unrepresentative-data.html}{IBM AI Risk Atlas - Unrepresentative data}\newline
\end{definitionbox}
\begin{definitionbox}{AI agents' impact on human agency}
The autonomous nature of AI agents in performing tasks or taking actions could affect the individuals' ability to engage in critical thinking, make choices and act independently.\newline\newline
\textbf{Concern: }AI agents might shift the decision, thinking, and control from humans to machines.  This might negatively impact the society and human welfare as they limit the freedom and meaningful participations of humans in performing a task or making decisions. \newline\newline
\textbf{Type: }agentic\newline
\textbf{Descriptor: }amplified by agentic AI \newline\newline
\textbf{Implementation details: } \newline
ID: atlas-impact-human-agency \newline
Tag: impact-human-agency \newline
URI:  \href{https://www.ibm.com/docs/en/watsonx/saas?topic=SSYOK8/wsj/ai-risk-atlas/impact-human-agency.html}{IBM AI Risk Atlas - AI agents' impact on human agency}\newline
\end{definitionbox}
\begin{definitionbox}{Personal information in prompt}
Personal information or sensitive personal information that is included as a part of a prompt that is sent to the model.\newline\newline
\textbf{Concern: }If personal information or sensitive personal information is included in the prompt, it might be unintentionally disclosed in the models' output. In addition to accidental disclosure, prompt data might be stored or later used for other purposes like model evaluation and retraining, and might appear in their output if not properly removed. \newline\newline
\textbf{Type: }inference\newline
\textbf{Descriptor: }specific to generative AI \newline\newline
\textbf{Implementation details: } \newline
ID: atlas-personal-information-in-prompt \newline
Tag: personal-information-in-prompt \newline
URI:  \href{https://www.ibm.com/docs/en/watsonx/saas?topic=SSYOK8/wsj/ai-risk-atlas/personal-information-in-prompt.html}{IBM AI Risk Atlas - Personal information in prompt}\newline
\end{definitionbox}
\begin{definitionbox}{AI agents' Impact on human agency}
The autonomous nature of AI agents in performing tasks or taking actions might affect the individuals' ability to engage in critical thinking, make choices, and acting independently.\newline\newline
\textbf{Concern: }AI agents might shift the decision, thinking, and control from humans to machines.  This might negatively impact society and human welfare as they limit the freedom and meaningful participations of humans in performing a task or making decisions.\newline\newline
\textbf{Type: }non-technical\newline
\textbf{Descriptor: }amplified by generative AI \newline\newline
\textbf{Implementation details: } \newline
ID: atlas-impact-on-human-agency \newline
Tag: impact-on-human-agency \newline
URI:  \href{https://www.ibm.com/docs/en/watsonx/saas?topic=SSYOK8/wsj/ai-risk-atlas/impact-on-human-agency.html}{IBM AI Risk Atlas - AI agents' Impact on human agency}\newline
\end{definitionbox}
\begin{definitionbox}{Sharing IP/PI/confidential information with user}
AI agents with unrestricted access to resources or databases or tools could potentially store and share PI/IP/confidential information with system users when performing their actions.\newline\newline
\textbf{Concern: }AI agents may share privileged information to users. The act of sharing the information may result in harm for the model owner, user, or others. The harm can vary based on the type and details of the information shared. Without adequate oversight, these privacy incidents might overwhelm company resources.\newline\newline
\textbf{Type: }agentic\newline
\textbf{Descriptor: }amplified by agentic AI \newline\newline
\textbf{Implementation details: } \newline
ID: atlas-sharing-info-user \newline
Tag: sharing-info-user \newline
URI:  \href{https://www.ibm.com/docs/en/watsonx/saas?topic=SSYOK8/wsj/ai-risk-atlas/sharing-info-user.html}{IBM AI Risk Atlas - Sharing IP/PI/confidential information with user}\newline
\end{definitionbox}
\begin{definitionbox}{Lack of testing diversity}
AI model risks are socio-technical, so their testing needs input from a broad set of disciplines and diverse testing practices.\newline\newline
\textbf{Concern: }Without diversity and the relevant experience, an organization might not correctly or completely identify and test for AI risks.\newline\newline
\textbf{Type: }non-technical\newline
\textbf{Descriptor: }amplified by generative AI \newline\newline
\textbf{Implementation details: } \newline
ID: atlas-lack-of-testing-diversity \newline
Tag: lack-of-testing-diversity \newline
URI:  \href{https://www.ibm.com/docs/en/watsonx/saas?topic=SSYOK8/wsj/ai-risk-atlas/lack-of-testing-diversity.html}{IBM AI Risk Atlas - Lack of testing diversity}\newline
\end{definitionbox}
\begin{definitionbox}{Nonconsensual use}
Generative AI models might be intentionally used to imitate people through deepfakes by using video, images, audio, or other modalities without their consent.\newline\newline
\textbf{Concern: }Deepfakes can spread disinformation about a person, possibly resulting in a negative impact on the person's reputation. A model that has this potential must be properly governed.\newline\newline
\textbf{Type: }output\newline
\textbf{Descriptor: }specific to generative AI \newline\newline
\textbf{Implementation details: } \newline
ID: atlas-nonconsensual-use \newline
Tag: nonconsensual-use \newline
URI:  \href{https://www.ibm.com/docs/en/watsonx/saas?topic=SSYOK8/wsj/ai-risk-atlas/nonconsensual-use.html}{IBM AI Risk Atlas - Nonconsensual use}\newline
\end{definitionbox}
\begin{definitionbox}{Decision bias}
Decision bias occurs when one group is unfairly advantaged over another due to decisions of the model. This might be caused by biases in the data and also amplified as a result of the model's training.\newline\newline
\textbf{Concern: }Bias can harm persons affected by the decisions of the model.\newline\newline
\textbf{Type: }output\newline
\textbf{Descriptor: }traditional risk of AI \newline\newline
\textbf{Implementation details: } \newline
ID: atlas-decision-bias \newline
Tag: decision-bias \newline
URI:  \href{https://www.ibm.com/docs/en/watsonx/saas?topic=SSYOK8/wsj/ai-risk-atlas/decision-bias.html}{IBM AI Risk Atlas - Decision bias}\newline
\end{definitionbox}
\begin{definitionbox}{Exposing personal information}
When personal identifiable information (PII) or sensitive personal information (SPI) are used in training data, fine-tuning data, or as part of the prompt, models might reveal that data in the generated output. Revealing personal information is a type of data leakage.\newline\newline
\textbf{Concern: }Sharing people's PI impacts their rights and make them more vulnerable.\newline\newline
\textbf{Type: }output\newline
\textbf{Descriptor: }amplified by generative AI \newline\newline
\textbf{Implementation details: } \newline
ID: atlas-exposing-personal-information \newline
Tag: exposing-personal-information \newline
URI:  \href{https://www.ibm.com/docs/en/watsonx/saas?topic=SSYOK8/wsj/ai-risk-atlas/exposing-personal-information.html}{IBM AI Risk Atlas - Exposing personal information}\newline
\end{definitionbox}
\begin{definitionbox}{AI agents' impact on jobs}
Widespread adoption of AI agents to perform complex tasks might lead to widespread automation of roles and could lead to job displacement.\newline\newline
\textbf{Concern: }As trust in agentic systems increases, business may be more motivated to use agents instead of people. Job displacement might lead to a loss of income and thus might negatively impact society and human welfare. Re-skilling may be challenging given the pace of the technology evolution.\newline\newline
\textbf{Type: }agentic\newline
\textbf{Descriptor: }amplified by agentic AI \newline\newline
\textbf{Implementation details: } \newline
ID: atlas-impact-jobs \newline
Tag: impact-jobs \newline
URI:  \href{https://www.ibm.com/docs/en/watsonx/saas?topic=SSYOK8/wsj/ai-risk-atlas/impact-jobs.html}{IBM AI Risk Atlas - AI agents' impact on jobs}\newline
\end{definitionbox}
\begin{definitionbox}{Improper data curation}
Improper collection and preparation of training or tuning data includes data label errors and by using data with conflicting information or misinformation.\newline\newline
\textbf{Concern: }Improper data curation can adversely affect how a model is trained, resulting in a model that does not behave in accordance with the intended values. Correcting problems after the model is trained and deployed might be insufficient for guaranteeing proper behavior. \newline\newline
\textbf{Type: }training-data\newline
\textbf{Descriptor: }amplified by generative AI \newline\newline
\textbf{Implementation details: } \newline
ID: atlas-data-curation \newline
Tag: data-curation \newline
URI:  \href{https://www.ibm.com/docs/en/watsonx/saas?topic=SSYOK8/wsj/ai-risk-atlas/data-curation.html}{IBM AI Risk Atlas - Improper data curation}\newline
\end{definitionbox}
\begin{definitionbox}{Over- or under-reliance on AI agents}
Reliance, that is the willingness to accept an AI agent behavior, depends on how much a user trusts that agent and what they are using it for. Over-reliance occurs when a user puts too much trust in an AI agent, accepting an AI agent's behavior even when it is likely undesired. Under-reliance is the opposite, where the user doesn't trust the AI agent but should. Increasing autonomy (to take action, select and consult resources/tools) of AI agents and the possibility of opaqueness and open-endedness increase the variability and visibility of agent behavior leading to difficulty in calibrating trust and possibly contributing to both over- and under-reliance.\newline\newline
\textbf{Concern: }Over/under reliance can lead to poor decision making by humans because of their misplaced trust in the AI agent, with negative consequences that escalate with the significance of the decision.\newline\newline
\textbf{Type: }agentic\newline
\textbf{Descriptor: }amplified by agentic AI \newline\newline
\textbf{Implementation details: } \newline
ID: atlas-over-or-under-reliance-on-ai-agents \newline
Tag: over-or-under-reliance-on-ai-agents \newline
URI:  \href{https://www.ibm.com/docs/en/watsonx/saas?topic=SSYOK8/wsj/ai-risk-atlas/over-or-under-reliance-on-ai-agents.html}{IBM AI Risk Atlas - Over- or under-reliance on AI agents}\newline
\end{definitionbox}
\begin{definitionbox}{Attack on AI agents’ external resources}
Attackers intentionally create vulnerabilities or exploit existing vulnerabilities in external resources (tools/database/applications/services/other agents) that AI agents rely on to execute their intended actions or to achieve their goals. \newline\newline
\textbf{Concern: }Compromised external resources could impact the AI agent's performance in different ways, such as manipulating AI agents to pursue a different goal, manipulating AI agents to execute undesired actions, capturing and relaying interactions between AI agents to malicious actors, and getting AI agents to share personal or confidential information.\newline\newline
\textbf{Type: }agentic\newline
\textbf{Descriptor: }specific to agentic AI \newline\newline
\textbf{Implementation details: } \newline
ID: atlas-external-resources-attack \newline
Tag: external-resources-attack \newline
URI:  \href{https://www.ibm.com/docs/en/watsonx/saas?topic=SSYOK8/wsj/ai-risk-atlas/external-resources-attack.html}{IBM AI Risk Atlas - Attack on AI agents’ external resources}\newline
\end{definitionbox}
\begin{definitionbox}{Revealing confidential information}
When confidential information is used in training data, fine-tuning data, or as part of the prompt, models might reveal that data in the generated output. Revealing confidential information is a type of data leakage.\newline\newline
\textbf{Concern: }If not properly developed to secure confidential data, the model might reveal confidential information or IP in the generated output and reveal information that was meant to be secret.\newline\newline
\textbf{Type: }output\newline
\textbf{Descriptor: }amplified by generative AI \newline\newline
\textbf{Implementation details: } \newline
ID: atlas-revealing-confidential-information \newline
Tag: revealing-confidential-information \newline
URI:  \href{https://www.ibm.com/docs/en/watsonx/saas?topic=SSYOK8/wsj/ai-risk-atlas/revealing-confidential-information.html}{IBM AI Risk Atlas - Revealing confidential information}\newline
\end{definitionbox}
\begin{definitionbox}{Spreading disinformation}
Generative AI models might be used to intentionally create misleading or false information to deceive or influence a targeted audience.\newline\newline
\textbf{Concern: }Spreading disinformation might affect human's ability to make informed decisions. A model that has this potential must be properly governed.\newline\newline
\textbf{Type: }output\newline
\textbf{Descriptor: }specific to generative AI \newline\newline
\textbf{Implementation details: } \newline
ID: atlas-spreading-disinformation \newline
Tag: spreading-disinformation \newline
URI:  \href{https://www.ibm.com/docs/en/watsonx/saas?topic=SSYOK8/wsj/ai-risk-atlas/spreading-disinformation.html}{IBM AI Risk Atlas - Spreading disinformation}\newline
\end{definitionbox}
\begin{definitionbox}{Uncertain data provenance}
Data provenance refers to tracing history of data, which includes its ownership, origin, and transformations. Without standardized and established methods for verifying where the data came from, there are no guarantees that the data is the same as the original source and has the correct usage terms.\newline\newline
\textbf{Concern: }Not all data sources are trustworthy. Data might be unethically collected, manipulated, or falsified. Verifying that data provenance is challenging due to factors such as data volume, data complexity, data source varieties, and poor data management. Using such data can result in undesirable behaviors in the model.\newline\newline
\textbf{Type: }training-data\newline
\textbf{Descriptor: }amplified by generative AI \newline\newline
\textbf{Implementation details: } \newline
ID: atlas-data-provenance \newline
Tag: data-provenance \newline
URI:  \href{https://www.ibm.com/docs/en/watsonx/saas?topic=SSYOK8/wsj/ai-risk-atlas/data-provenance.html}{IBM AI Risk Atlas - Uncertain data provenance}\newline
\end{definitionbox}
\begin{definitionbox}{Unrepresentative risk testing}
Testing is unrepresentative when the test inputs are mismatched with the inputs that are expected during deployment.\newline\newline
\textbf{Concern: }If the model is evaluated in a use, context, or setting that is not the same as the one expected for deployment, the evaluations might not accurately reflect the risks of the model.\newline\newline
\textbf{Type: }non-technical\newline
\textbf{Descriptor: }amplified by generative AI \newline\newline
\textbf{Implementation details: } \newline
ID: atlas-unrepresentative-risk-testing \newline
Tag: unrepresentative-risk-testing \newline
URI:  \href{https://www.ibm.com/docs/en/watsonx/saas?topic=SSYOK8/wsj/ai-risk-atlas/unrepresentative-risk-testing.html}{IBM AI Risk Atlas - Unrepresentative risk testing}\newline
\end{definitionbox}
\begin{definitionbox}{Data bias}
Historical and societal biases that are present in the data are used to train and fine-tune the model.\newline\newline
\textbf{Concern: }Training an AI system on data with bias, such as historical or societal bias, can lead to biased or skewed outputs that can unfairly represent or otherwise discriminate against certain groups or individuals.\newline\newline
\textbf{Type: }training-data\newline
\textbf{Descriptor: }amplified by generative AI \newline\newline
\textbf{Implementation details: } \newline
ID: atlas-data-bias \newline
Tag: data-bias \newline
URI:  \href{https://www.ibm.com/docs/en/watsonx/saas?topic=SSYOK8/wsj/ai-risk-atlas/data-bias.html}{IBM AI Risk Atlas - Data bias}\newline
\end{definitionbox}
\begin{definitionbox}{Data usage rights restrictions}
Terms of service, license compliance, or other IP issues may restrict the ability to use certain data for building models.\newline\newline
\textbf{Concern: }Laws and regulations concerning the use of data to train AI are unsettled and can vary from country to country, which creates challenges in the development of models.\newline\newline
\textbf{Type: }training-data\newline
\textbf{Descriptor: }amplified by generative AI \newline\newline
\textbf{Implementation details: } \newline
ID: atlas-data-usage-rights \newline
Tag: data-usage-rights \newline
URI:  \href{https://www.ibm.com/docs/en/watsonx/saas?topic=SSYOK8/wsj/ai-risk-atlas/data-usage-rights.html}{IBM AI Risk Atlas - Data usage rights restrictions}\newline
\end{definitionbox}
\begin{definitionbox}{Unauthorized use}
Unauthorized use: If attackers can gain access to the AI agent and its components, they can perform actions that can have different levels of harm depending on the agent's capabilities and information it has access to. Examples: Using stored personal information to mimic identity or impersonate with an intent to deceive. Manipulating AI agent's behavior via feedback to the AI agent or corrupting its memory to change its behavior. Manipulating the problem description or the goal to get the AI agent to behave badly or run harmful commands .\newline\newline
\textbf{Concern: }Attackers accessing the agent can alter AI agent's behavior and make it execute actions that benefit the attacker such as executing actions that lead to system degradation, data exfiltration, exhausting available resources, and impairing performance. The actions taken by the attackers may cause harms to others.\newline\newline
\textbf{Type: }agentic\newline
\textbf{Descriptor: }amplified by agentic AI \newline\newline
\textbf{Implementation details: } \newline
ID: atlas-unauthorized-use \newline
Tag: unauthorized-use \newline
URI:  \href{https://www.ibm.com/docs/en/watsonx/saas?topic=SSYOK8/wsj/ai-risk-atlas/unauthorized-use.html}{IBM AI Risk Atlas - Unauthorized use}\newline
\end{definitionbox}
\begin{definitionbox}{Redundant actions}
AI agents can execute actions that are not needed for achieving the goal. In an extreme case, AI agents might enter a cycle of executing the same actions repeatedly without any progress. This could happen because of unexpected conditions in the environment, the AI agent's failure to reflect on its action, AI agent reasoning and planning errors or the AI agent's lack of knowledge about the problem.\newline\newline
\textbf{Concern: }Executing actions that are not needed for the goal might result in wasting computation resources, increased cost, reducing AI agent's efficiency in achieving the goal, and leading to potentially harmful outcomes. Executing the same actions repeatedly could prevent the AI agent from achieving the goal, strain computational resources, and increase cost. As agents become more autonomous, verifying that AI agents operate efficiently becomes increasing time consuming.\newline\newline
\textbf{Type: }agentic\newline
\textbf{Descriptor: }specific to agentic AI \newline\newline
\textbf{Implementation details: } \newline
ID: atlas-redundant-actions \newline
Tag: redundant-actions \newline
URI:  \href{https://www.ibm.com/docs/en/watsonx/saas?topic=SSYOK8/wsj/ai-risk-atlas/redundant-actions.html}{IBM AI Risk Atlas - Redundant actions}\newline
\end{definitionbox}
\begin{definitionbox}{AI agents' impact on environment}
Complexity of the tasks and possibility of AI agents performing redundant actions could lead to computational inefficiencies and add to the environmental impact.\newline\newline
\textbf{Concern: }The operation of AI agents could contribute to carbon emissions. If not managed, these could exacerbate climate change.\newline\newline
\textbf{Type: }agentic\newline
\textbf{Descriptor: }amplified by agentic AI \newline\newline
\textbf{Implementation details: } \newline
ID: atlas-impact-environment \newline
Tag: impact-environment \newline
URI:  \href{https://www.ibm.com/docs/en/watsonx/saas?topic=SSYOK8/wsj/ai-risk-atlas/impact-environment.html}{IBM AI Risk Atlas - AI agents' impact on environment}\newline
\end{definitionbox}
\begin{definitionbox}{Misaligned actions}
AI agents can take actions that are not aligned with relevant human values, ethical considerations, guidelines and policies. Misaligned actions can occur in different ways such as: Applying learned goals inappropriately to new or unforeseen situations. Using AI agents for a purpose/goals that are beyond their intended use. Selecting resources or tools in a biased way. Using deceptive tactics to achieve the goal by developing the capacity for scheming based on the instructions given within a specific context. Compromising on AI agent values to work with another AI agent or tool to accomplish the task. \newline\newline
\textbf{Concern: }Misaligned actions can adversely impact or harm people. \newline\newline
\textbf{Type: }agentic\newline
\textbf{Descriptor: }amplified by agentic AI \newline\newline
\textbf{Implementation details: } \newline
ID: atlas-misaligned-actions \newline
Tag: misaligned-actions \newline
URI:  \href{https://www.ibm.com/docs/en/watsonx/saas?topic=SSYOK8/wsj/ai-risk-atlas/misaligned-actions.html}{IBM AI Risk Atlas - Misaligned actions}\newline
\end{definitionbox}
\begin{definitionbox}{Data contamination}
Data contamination occurs when incorrect data is used for training. For example, data that is not aligned with model's purpose or data that is already set aside for other development tasks such as testing and evaluation.\newline\newline
\textbf{Concern: }Data that differs from the intended training data might skew model accuracy and affect model outcomes.\newline\newline
\textbf{Type: }training-data\newline
\textbf{Descriptor: }amplified by generative AI \newline\newline
\textbf{Implementation details: } \newline
ID: atlas-data-contamination \newline
Tag: data-contamination \newline
URI:  \href{https://www.ibm.com/docs/en/watsonx/saas?topic=SSYOK8/wsj/ai-risk-atlas/data-contamination.html}{IBM AI Risk Atlas - Data contamination}\newline
\end{definitionbox}
\begin{definitionbox}{Harmful code generation}
Models might generate code that causes harm or unintentionally affects other systems.\newline\newline
\textbf{Concern: }The execution of harmful code might open vulnerabilities in IT systems.\newline\newline
\textbf{Type: }output\newline
\textbf{Descriptor: }specific to generative AI \newline\newline
\textbf{Implementation details: } \newline
ID: atlas-harmful-code-generation \newline
Tag: harmful-code-generation \newline
URI:  \href{https://www.ibm.com/docs/en/watsonx/saas?topic=SSYOK8/wsj/ai-risk-atlas/harmful-code-generation.html}{IBM AI Risk Atlas - Harmful code generation}\newline
\end{definitionbox}
\begin{definitionbox}{Incomplete usage definition}
Since foundation models can be used for many purposes, a model's intended use is important for defining the relevant risks of that model. As the use changes, the relevant risks might correspondingly change.\newline\newline
\textbf{Concern: }It might be difficult to accurately determine and mitigate the relevant risks for a model when its intended use is insufficiently specified. Such as how a model is going to be used, where it is going to be used and what it is going to be used for.\newline\newline
\textbf{Type: }non-technical\newline
\textbf{Descriptor: }specific to generative AI \newline\newline
\textbf{Implementation details: } \newline
ID: atlas-incomplete-usage-definition \newline
Tag: incomplete-usage-definition \newline
URI:  \href{https://www.ibm.com/docs/en/watsonx/saas?topic=SSYOK8/wsj/ai-risk-atlas/incomplete-usage-definition.html}{IBM AI Risk Atlas - Incomplete usage definition}\newline
\end{definitionbox}
\begin{definitionbox}{Lack of data transparency}
Lack of data transparency is due to insufficient documentation of training or tuning dataset details. \newline\newline
\textbf{Concern: }Transparency is important for legal compliance and AI ethics. Information on the collection and preparation of training data, including how it was labeled and by who are necessary to understand model behavior and suitability. Details about how the data risks were determined, measured, and mitigated are important for evaluating both data and model trustworthiness. Missing details about the data might make it more difficult to evaluate representational harms, data ownership, provenance, and other data-oriented risks. The lack of standardized requirements might limit disclosure as organizations protect trade secrets and try to limit others from copying their models.\newline\newline
\textbf{Type: }non-technical\newline
\textbf{Descriptor: }amplified by generative AI \newline\newline
\textbf{Implementation details: } \newline
ID: atlas-lack-of-data-transparency \newline
Tag: lack-of-data-transparency \newline
URI:  \href{https://www.ibm.com/docs/en/watsonx/saas?topic=SSYOK8/wsj/ai-risk-atlas/lack-of-data-transparency.html}{IBM AI Risk Atlas - Lack of data transparency}\newline
\end{definitionbox}
\begin{definitionbox}{Copyright infringement}
A model might generate content that is similar or identical to existing work protected by copyright or covered by open-source license agreement.\newline\newline
\textbf{Concern: }Laws and regulations concerning the use of content that looks the same or closely similar to other copyrighted data are largely unsettled and can vary from country to country, providing challenges in determining and implementing compliance.\newline\newline
\textbf{Type: }output\newline
\textbf{Descriptor: }specific to generative AI \newline\newline
\textbf{Implementation details: } \newline
ID: atlas-copyright-infringement \newline
Tag: copyright-infringement \newline
URI:  \href{https://www.ibm.com/docs/en/watsonx/saas?topic=SSYOK8/wsj/ai-risk-atlas/copyright-infringement.html}{IBM AI Risk Atlas - Copyright infringement}\newline
\end{definitionbox}
\begin{definitionbox}{Context overload attack}
Overloading the prompt with excessive tokens, for instance with many-shot examples, can predispose models to a vulnerable state.\newline\newline
\textbf{Concern: }Context overload attacks can be used to alter model behavior and benefit the attacker. The content it generates may cause harms for the user or others.\newline\newline
\textbf{Type: }inference\newline
\textbf{Descriptor: }specific to generative AI \newline\newline
\textbf{Implementation details: } \newline
ID: atlas-context-overload-attack \newline
Tag: context-overload-attack \newline
URI:  \href{https://www.ibm.com/docs/en/watsonx/saas?topic=SSYOK8/wsj/ai-risk-atlas/context-overload-attack.html}{IBM AI Risk Atlas - Context overload attack}\newline
\end{definitionbox}
\begin{definitionbox}{Impact on affected communities}
It is important to include the perspectives or concerns of communities that are affected by model outcomes when designing and building models. Failing to include these perspectives makes it difficult to understand the relevant context for the model and to engender trust within these communities.\newline\newline
\textbf{Concern: }Failing to engage with communities that are affected by a model's outcomes might result in harms to those communities and societal backlash.\newline\newline
\textbf{Type: }non-technical\newline
\textbf{Descriptor: }traditional risk of AI \newline\newline
\textbf{Implementation details: } \newline
ID: atlas-impact-on-affected-communities \newline
Tag: impact-on-affected-communities \newline
URI:  \href{https://www.ibm.com/docs/en/watsonx/saas?topic=SSYOK8/wsj/ai-risk-atlas/impact-on-affected-communities.html}{IBM AI Risk Atlas - Impact on affected communities}\newline
\end{definitionbox}
\begin{definitionbox}{Improper retraining}
Using undesirable output (for example, inaccurate, inappropriate, and user content) for retraining purposes can result in unexpected model behavior.\newline\newline
\textbf{Concern: }Repurposing generated output for retraining a model without implementing proper human vetting increases the chances of undesirable outputs to be incorporated into the training or tuning data of the model. In turn, this model can generate even more undesirable output.\newline\newline
\textbf{Type: }training-data\newline
\textbf{Descriptor: }amplified by generative AI \newline\newline
\textbf{Implementation details: } \newline
ID: atlas-improper-retraining \newline
Tag: improper-retraining \newline
URI:  \href{https://www.ibm.com/docs/en/watsonx/saas?topic=SSYOK8/wsj/ai-risk-atlas/improper-retraining.html}{IBM AI Risk Atlas - Improper retraining}\newline
\end{definitionbox}
\begin{definitionbox}{Spreading toxicity}
Generative AI models might be used intentionally to generate hateful, abusive, and profane (HAP) or obscene content.\newline\newline
\textbf{Concern: }Toxic content might negatively affect the well-being of its recipients. A model that has this potential must be properly governed.\newline\newline
\textbf{Type: }output\newline
\textbf{Descriptor: }specific to generative AI \newline\newline
\textbf{Implementation details: } \newline
ID: atlas-spreading-toxicity \newline
Tag: spreading-toxicity \newline
URI:  \href{https://www.ibm.com/docs/en/watsonx/saas?topic=SSYOK8/wsj/ai-risk-atlas/spreading-toxicity.html}{IBM AI Risk Atlas - Spreading toxicity}\newline
\end{definitionbox}
\begin{definitionbox}{Introduce data bias}
Specific actions taken by the AI agent, such as modifying a dataset or a database, can introduce bias in the resource that gets used by others or by itself to take actions.\newline\newline
\textbf{Concern: }AI agents can introduce or magnify existing discriminatory behaviors. It can harm people depending on the use.\newline\newline
\textbf{Type: }agentic\newline
\textbf{Descriptor: }amplified by agentic AI \newline\newline
\textbf{Implementation details: } \newline
ID: atlas-introduce-data-bias \newline
Tag: introduce-data-bias \newline
URI:  \href{https://www.ibm.com/docs/en/watsonx/saas?topic=SSYOK8/wsj/ai-risk-atlas/introduce-data-bias.html}{IBM AI Risk Atlas - Introduce data bias}\newline
\end{definitionbox}
\begin{definitionbox}{Accountability of AI agent actions}
Assigning responsibility for an action taken by an agentic AI system is difficult due to the complexity of agents and the number of external resources, tools or agents they interact with.\newline\newline
\textbf{Concern: }Without properly documenting decisions and assigning responsibility, determining liability for unexpected behavior or misuse might not be possible.\newline\newline
\textbf{Type: }agentic\newline
\textbf{Descriptor: }amplified by agentic AI \newline\newline
\textbf{Implementation details: } \newline
ID: atlas-accountability \newline
Tag: accountability \newline
URI:  \href{https://www.ibm.com/docs/en/watsonx/saas?topic=SSYOK8/wsj/ai-risk-atlas/accountability.html}{IBM AI Risk Atlas - Accountability of AI agent actions}\newline
\end{definitionbox}
\begin{definitionbox}{Incomplete AI agent evaluation}
Evaluating the performance or accuracy or an agent is difficult because of system complexity and open-endedness.\newline\newline
\textbf{Concern: }Insufficient evaluation of an agent's performance or accuracy can lead to the use of agents that do not perform to expectations. Incorrect agent behavior can result in harms to an agent's users or others.\newline\newline
\textbf{Type: }agentic\newline
\textbf{Descriptor: }amplified by agentic AI \newline\newline
\textbf{Implementation details: } \newline
ID: atlas-incomplete-ai-agent-evaluation \newline
Tag: incomplete-ai-agent-evaluation \newline
URI:  \href{https://www.ibm.com/docs/en/watsonx/saas?topic=SSYOK8/wsj/ai-risk-atlas/incomplete-ai-agent-evaluation.html}{IBM AI Risk Atlas - Incomplete AI agent evaluation}\newline
\end{definitionbox}
\begin{definitionbox}{Inaccessible training data}
Without access to the training data, the types of explanations a model can provide are limited and more likely to be incorrect.\newline\newline
\textbf{Concern: }Low quality explanations without source data make it difficult for users, model validators, and auditors to understand and trust the model.\newline\newline
\textbf{Type: }output\newline
\textbf{Descriptor: }amplified by generative AI \newline\newline
\textbf{Implementation details: } \newline
ID: atlas-inaccessible-training-data \newline
Tag: inaccessible-training-data \newline
URI:  \href{https://www.ibm.com/docs/en/watsonx/saas?topic=SSYOK8/wsj/ai-risk-atlas/inaccessible-training-data.html}{IBM AI Risk Atlas - Inaccessible training data}\newline
\end{definitionbox}
\begin{definitionbox}{Impact on education: bypassing learning}
Easy access to high-quality generative models might result in students that use AI models to bypass the learning process.\newline\newline
\textbf{Concern: }AI models are quick to find solutions or solve complex problems. These systems can be misused by students to bypass the learning process. The ease of access to these models results in students having a superficial understanding of concepts and hampers further education that might rely on understanding those concepts.\newline\newline
\textbf{Type: }non-technical\newline
\textbf{Descriptor: }specific to generative AI \newline\newline
\textbf{Implementation details: } \newline
ID: atlas-bypassing-learning \newline
Tag: bypassing-learning \newline
URI:  \href{https://www.ibm.com/docs/en/watsonx/saas?topic=SSYOK8/wsj/ai-risk-atlas/bypassing-learning.html}{IBM AI Risk Atlas - Impact on education: bypassing learning}\newline
\end{definitionbox}
\begin{definitionbox}{Untraceable attribution}
The content of the training data used for generating the model's output is not accessible.\newline\newline
\textbf{Concern: }Without the ability to access training data content, the possibility of using source attribution techniques can be severely limited or impossible. This makes it difficult for users, model validators, and auditors to understand and trust the model.\newline\newline
\textbf{Type: }output\newline
\textbf{Descriptor: }amplified by generative AI \newline\newline
\textbf{Implementation details: } \newline
ID: atlas-untraceable-attribution \newline
Tag: untraceable-attribution \newline
URI:  \href{https://www.ibm.com/docs/en/watsonx/saas?topic=SSYOK8/wsj/ai-risk-atlas/untraceable-attribution.html}{IBM AI Risk Atlas - Untraceable attribution}\newline
\end{definitionbox}
\begin{definitionbox}{Non-disclosure}
Content might not be clearly disclosed as AI generated.\newline\newline
\textbf{Concern: }Users must be notified when they are interacting with an AI system. Not disclosing the AI-authored content can result in a lack of transparency.\newline\newline
\textbf{Type: }output\newline
\textbf{Descriptor: }specific to generative AI \newline\newline
\textbf{Implementation details: } \newline
ID: atlas-non-disclosure \newline
Tag: non-disclosure \newline
URI:  \href{https://www.ibm.com/docs/en/watsonx/saas?topic=SSYOK8/wsj/ai-risk-atlas/non-disclosure.html}{IBM AI Risk Atlas - Non-disclosure}\newline
\end{definitionbox}
\begin{definitionbox}{Lack of training data transparency}
Without accurate documentation on how a model's data was collected, curated, and used to train a model, it might be harder to satisfactorily explain the behavior of the model with respect to the data.\newline\newline
\textbf{Concern: }A lack of data documentation limits the ability to evaluate risks associated with the data. Having access to the training data is not enough. Without recording how the data was cleaned, modified, or generated, the model behavior is more difficult to understand and to fix. Lack of data transparency also impacts model reuse as it is difficult to determine data representativeness for the new use without such documentation.\newline\newline
\textbf{Type: }training-data\newline
\textbf{Descriptor: }amplified by generative AI \newline\newline
\textbf{Implementation details: } \newline
ID: atlas-data-transparency \newline
Tag: data-transparency \newline
URI:  \href{https://www.ibm.com/docs/en/watsonx/saas?topic=SSYOK8/wsj/ai-risk-atlas/data-transparency.html}{IBM AI Risk Atlas - Lack of training data transparency}\newline
\end{definitionbox}
\begin{definitionbox}{Model usage rights restrictions}
Terms of service, licenses, or other rules restrict the use of certain models.\newline\newline
\textbf{Concern: }Laws and regulations that concern the use of AI are in place and vary from country to country. Additionally, the usage of models might be dictated by licensing terms or agreements.\newline\newline
\textbf{Type: }non-technical\newline
\textbf{Descriptor: }traditional risk of AI \newline\newline
\textbf{Implementation details: } \newline
ID: atlas-model-usage-rights \newline
Tag: model-usage-rights \newline
URI:  \href{https://www.ibm.com/docs/en/watsonx/saas?topic=SSYOK8/wsj/ai-risk-atlas/model-usage-rights.html}{IBM AI Risk Atlas - Model usage rights restrictions}\newline
\end{definitionbox}
\begin{definitionbox}{Reproducibility}
Replicating agent behavior or output can be impacted by changes or updates made to external services and tools. This impact is increased if the agent is built with generative AI.\newline\newline
\textbf{Concern: }Because AI agents behavior may rely on Application Programming Interfaces (APIs), systems, or other resources that may change or become unavailable, evaluations that rely on reproducible results may not be reliably reproduced. This adds cost and complexity to the development and evaluation of agents. Not being able to reproduce results could impact reliance of humans on the AI agents.\newline\newline
\textbf{Type: }agentic\newline
\textbf{Descriptor: }specific to agentic AI \newline\newline
\textbf{Implementation details: } \newline
ID: atlas-reproducibility \newline
Tag: reproducibility \newline
URI:  \href{https://www.ibm.com/docs/en/watsonx/saas?topic=SSYOK8/wsj/ai-risk-atlas/reproducibility.html}{IBM AI Risk Atlas - Reproducibility}\newline
\end{definitionbox}
\begin{definitionbox}{Specialized tokens attack}
Prompt attacks that include specialized tokens, often algorithmically designed, to target and exploit vulnerabilities in the model.\newline\newline
\textbf{Concern: }Specialized tokens attacks can be used to alter model behavior and benefit the attacker. The content it generates may cause harms for the user or others.\newline\newline
\textbf{Type: }inference\newline
\textbf{Descriptor: }specific to generative AI \newline\newline
\textbf{Implementation details: } \newline
ID: atlas-specialized-tokens-attack \newline
Tag: specialized-tokens-attack \newline
URI:  \href{https://www.ibm.com/docs/en/watsonx/saas?topic=SSYOK8/wsj/ai-risk-atlas/specialized-tokens-attack.html}{IBM AI Risk Atlas - Specialized tokens attack}\newline
\end{definitionbox}
\begin{definitionbox}{Incomplete advice}
When a model provides advice without having enough information, resulting in possible harm if the advice is followed.\newline\newline
\textbf{Concern: }A person might act on incomplete advice or worry about a situation that is not applicable to them due to the overgeneralized nature of the content generated. For example, a model might provide incorrect medical, financial, and legal advice or recommendations that the end user might act on, resulting in harmful actions.\newline\newline
\textbf{Type: }output\newline
\textbf{Descriptor: }specific to generative AI \newline\newline
\textbf{Implementation details: } \newline
ID: atlas-incomplete-advice \newline
Tag: incomplete-advice \newline
URI:  \href{https://www.ibm.com/docs/en/watsonx/saas?topic=SSYOK8/wsj/ai-risk-atlas/incomplete-advice.html}{IBM AI Risk Atlas - Incomplete advice}\newline
\end{definitionbox}
\begin{definitionbox}{Prompt injection attack}
A prompt injection attack forces a generative model that takes a prompt as input to produce unexpected output by manipulating the structure, instructions or information contained in its prompt. Many types of prompt attacks exist as described in the prompt attack section of the table.\newline\newline
\textbf{Concern: }Injection attacks can be used to alter model behavior and benefit the attacker.\newline\newline
\textbf{Type: }inference\newline
\textbf{Descriptor: }specific to generative AI \newline\newline
\textbf{Implementation details: } \newline
ID: atlas-prompt-injection \newline
Tag: prompt-injection \newline
URI:  \href{https://www.ibm.com/docs/en/watsonx/saas?topic=SSYOK8/wsj/ai-risk-atlas/prompt-injection.html}{IBM AI Risk Atlas - Prompt injection attack}\newline
\end{definitionbox}
\begin{definitionbox}{Lack of system transparency}
Insufficient documentation of the system that uses the model and the model's purpose within the system in which it is used.\newline\newline
\textbf{Concern: }A lack of documentation makes it difficult to understand how the model's outcomes contribute to the system's or application's functionality.\newline\newline
\textbf{Type: }non-technical\newline
\textbf{Descriptor: }traditional risk of AI \newline\newline
\textbf{Implementation details: } \newline
ID: atlas-lack-of-system-transparency \newline
Tag: lack-of-system-transparency \newline
URI:  \href{https://www.ibm.com/docs/en/watsonx/saas?topic=SSYOK8/wsj/ai-risk-atlas/lack-of-system-transparency.html}{IBM AI Risk Atlas - Lack of system transparency}\newline
\end{definitionbox}
\begin{definitionbox}{Data usage restrictions}
Laws and other restrictions can limit or prohibit the use of some data for specific AI use cases.\newline\newline
\textbf{Concern: }Data usage restrictions can impact the availability of the data required for training an AI model and can lead to poorly represented data.\newline\newline
\textbf{Type: }training-data\newline
\textbf{Descriptor: }traditional risk of AI \newline\newline
\textbf{Implementation details: } \newline
ID: atlas-data-usage \newline
Tag: data-usage \newline
URI:  \href{https://www.ibm.com/docs/en/watsonx/saas?topic=SSYOK8/wsj/ai-risk-atlas/data-usage.html}{IBM AI Risk Atlas - Data usage restrictions}\newline
\end{definitionbox}
\begin{definitionbox}{Impact on cultural diversity}
AI systems might overly represent certain cultures that result in a homogenization of culture and thoughts.\newline\newline
\textbf{Concern: }Underrepresented groups' languages, viewpoints, and institutions might be suppressed by that means reducing diversity of thought and culture.\newline\newline
\textbf{Type: }non-technical\newline
\textbf{Descriptor: }specific to generative AI \newline\newline
\textbf{Implementation details: } \newline
ID: atlas-impact-on-cultural-diversity \newline
Tag: impact-on-cultural-diversity \newline
URI:  \href{https://www.ibm.com/docs/en/watsonx/saas?topic=SSYOK8/wsj/ai-risk-atlas/impact-on-cultural-diversity.html}{IBM AI Risk Atlas - Impact on cultural diversity}\newline
\end{definitionbox}
\begin{definitionbox}{Impact on education: plagiarism}
Easy access to high-quality generative models might result in students that use AI models to plagiarize existing work intentionally or unintentionally.\newline\newline
\textbf{Concern: }AI models can be used to claim the authorship or originality of works that were created by other people in doing so by engaging in plagiarism. Claiming others' work as your own is both unethical and often illegal.\newline\newline
\textbf{Type: }non-technical\newline
\textbf{Descriptor: }specific to generative AI \newline\newline
\textbf{Implementation details: } \newline
ID: atlas-plagiarism \newline
Tag: plagiarism \newline
URI:  \href{https://www.ibm.com/docs/en/watsonx/saas?topic=SSYOK8/wsj/ai-risk-atlas/plagiarism.html}{IBM AI Risk Atlas - Impact on education: plagiarism}\newline
\end{definitionbox}
\begin{definitionbox}{Personal information in data}
Inclusion or presence of personal identifiable information (PII) and sensitive personal information (SPI) in the data used for training or fine tuning the model might result in unwanted disclosure of that information.\newline\newline
\textbf{Concern: }If not properly developed to protect sensitive data, the model might expose personal information in the generated output.  Additionally, personal, or sensitive data must be reviewed and handled in accordance with privacy laws and regulations.\newline\newline
\textbf{Type: }training-data\newline
\textbf{Descriptor: }traditional risk of AI \newline\newline
\textbf{Implementation details: } \newline
ID: atlas-personal-information-in-data \newline
Tag: personal-information-in-data \newline
URI:  \href{https://www.ibm.com/docs/en/watsonx/saas?topic=SSYOK8/wsj/ai-risk-atlas/personal-information-in-data.html}{IBM AI Risk Atlas - Personal information in data}\newline
\end{definitionbox}
\begin{definitionbox}{Direct instructions attack}
Prompts, questions, or requests designed to elicit undesirable responses from the application. This approach directly instructs the model to engage in the undesired behavior.\newline\newline
\textbf{Concern: }Direct instructions attacks can be used to alter model behavior and benefit the attacker. The content it generates may cause harms for the user or others.\newline\newline
\textbf{Type: }inference\newline
\textbf{Descriptor: }specific to generative AI \newline\newline
\textbf{Implementation details: } \newline
ID: atlas-direct-instructions-attack \newline
Tag: direct-instructions-attack \newline
URI:  \href{https://www.ibm.com/docs/en/watsonx/saas?topic=SSYOK8/wsj/ai-risk-atlas/direct-instructions-attack.html}{IBM AI Risk Atlas - Direct instructions attack}\newline
\end{definitionbox}
\begin{definitionbox}{Improper usage}
Improper usage occurs when a model is used for a purpose that it was not originally designed for.\newline\newline
\textbf{Concern: }Reusing a model without understanding its original data, design intent, and goals might result in unexpected and unwanted model behaviors.\newline\newline
\textbf{Type: }output\newline
\textbf{Descriptor: }amplified by generative AI \newline\newline
\textbf{Implementation details: } \newline
ID: atlas-improper-usage \newline
Tag: improper-usage \newline
URI:  \href{https://www.ibm.com/docs/en/watsonx/saas?topic=SSYOK8/wsj/ai-risk-atlas/improper-usage.html}{IBM AI Risk Atlas - Improper usage}\newline
\end{definitionbox}
\begin{definitionbox}{Impact on Jobs}
Widespread adoption of foundation model-based AI systems might lead to people's job loss as their work is automated if they are not reskilled.\newline\newline
\textbf{Concern: }Job loss might lead to a loss of income and thus might negatively impact the society and human welfare. Reskilling might be challenging given the pace of the technology evolution.\newline\newline
\textbf{Type: }non-technical\newline
\textbf{Descriptor: }amplified by generative AI \newline\newline
\textbf{Implementation details: } \newline
ID: atlas-impact-on-jobs \newline
Tag: impact-on-jobs \newline
URI:  \href{https://www.ibm.com/docs/en/watsonx/saas?topic=SSYOK8/wsj/ai-risk-atlas/impact-on-jobs.html}{IBM AI Risk Atlas - Impact on Jobs}\newline
\end{definitionbox}
\begin{definitionbox}{Extraction attack}
An extraction attack attempts to copy or steal an AI model by appropriately sampling the input space and observing outputs to build a surrogate model that behaves similarly.\newline\newline
\textbf{Concern: }With a successful extraction attack, the attacker can perform further adversarial attacks to gain valuable information such as sensitive personal information or intellectual property.\newline\newline
\textbf{Type: }inference\newline
\textbf{Descriptor: }amplified by generative AI \newline\newline
\textbf{Implementation details: } \newline
ID: atlas-extraction-attack \newline
Tag: extraction-attack \newline
URI:  \href{https://www.ibm.com/docs/en/watsonx/saas?topic=SSYOK8/wsj/ai-risk-atlas/extraction-attack.html}{IBM AI Risk Atlas - Extraction attack}\newline
\end{definitionbox}
\begin{definitionbox}{Jailbreaking}
A jailbreaking attack attempts to break through the guardrails established in the model to perform restricted actions.\newline\newline
\textbf{Concern: }Jailbreaking attacks can be used to alter model behavior and benefit the attacker.\newline\newline
\textbf{Type: }inference\newline
\textbf{Descriptor: }specific to generative AI \newline\newline
\textbf{Implementation details: } \newline
ID: atlas-jailbreaking \newline
Tag: jailbreaking \newline
URI:  \href{https://www.ibm.com/docs/en/watsonx/saas?topic=SSYOK8/wsj/ai-risk-atlas/jailbreaking.html}{IBM AI Risk Atlas - Jailbreaking}\newline
\end{definitionbox}
\begin{definitionbox}{Data acquisition restrictions}
Laws and other regulations might limit the collection of certain types of data for specific AI use cases.\newline\newline
\textbf{Concern: }There are several ways of collecting data for building a foundation models: web scraping, web crawling, crowdsourcing, and curating public datasets. Data acquisition restrictions can also impact the availability of the data that is required for training an AI model and can lead to poorly represented data.\newline\newline
\textbf{Type: }training-data\newline
\textbf{Descriptor: }amplified by generative AI \newline\newline
\textbf{Implementation details: } \newline
ID: atlas-data-acquisition \newline
Tag: data-acquisition \newline
URI:  \href{https://www.ibm.com/docs/en/watsonx/saas?topic=SSYOK8/wsj/ai-risk-atlas/data-acquisition.html}{IBM AI Risk Atlas - Data acquisition restrictions}\newline
\end{definitionbox}
\begin{definitionbox}{Sharing IP/PI/confidential information with tools}
AI agents with unrestricted access to resources or databases or tools could potentially store and share PI/IP/confidential information with other tools or agents when performing their actions.\newline\newline
\textbf{Concern: }AI agents may share privileged information with other tools/agents. The act of sharing the information may result in harm for the model owner, user, or others. The harm can vary based on the type and details of the information shared. Without adequate oversight, these privacy incidents might overwhelm company resources.\newline\newline
\textbf{Type: }agentic\newline
\textbf{Descriptor: }specific to agentic AI \newline\newline
\textbf{Implementation details: } \newline
ID: atlas-sharing-info-tools \newline
Tag: sharing-info-tools \newline
URI:  \href{https://www.ibm.com/docs/en/watsonx/saas?topic=SSYOK8/wsj/ai-risk-atlas/sharing-info-tools.html}{IBM AI Risk Atlas - Sharing IP/PI/confidential information with tools}\newline
\end{definitionbox}
\begin{definitionbox}{Prompt priming}
Because generative models produce output based on the input provided, the model can be prompted to reveal specific kinds of information. For example, adding personal information in the prompt increases its likelihood of generating similar kinds of personal information in its output. If personal data was included as part of the model's training, there is a possibility it could be revealed.\newline\newline
\textbf{Concern: }The attack can be used to alter model behavior and benefit the attacker.\newline\newline
\textbf{Type: }inference\newline
\textbf{Descriptor: }specific to generative AI \newline\newline
\textbf{Implementation details: } \newline
ID: atlas-prompt-priming \newline
Tag: prompt-priming \newline
URI:  \href{https://www.ibm.com/docs/en/watsonx/saas?topic=SSYOK8/wsj/ai-risk-atlas/prompt-priming.html}{IBM AI Risk Atlas - Prompt priming}\newline
\end{definitionbox}
\begin{definitionbox}{Reidentification}
Even with the removal or personal identifiable information (PII) and sensitive personal information (SPI) from data, it might be possible to identify persons due to correlations to other features available in the data.\newline\newline
\textbf{Concern: }Including irrelevant but highly correlated features to personal information for model training can increase the risk of reidentification.\newline\newline
\textbf{Type: }training-data\newline
\textbf{Descriptor: }traditional risk of AI \newline\newline
\textbf{Implementation details: } \newline
ID: atlas-reidentification \newline
Tag: reidentification \newline
URI:  \href{https://www.ibm.com/docs/en/watsonx/saas?topic=SSYOK8/wsj/ai-risk-atlas/reidentification.html}{IBM AI Risk Atlas - Reidentification}\newline
\end{definitionbox}
\begin{definitionbox}{Attribute inference attack}
An attribute inference attack repeatedly queries a model to detect whether certain sensitive features can be inferred about individuals who participated in training a model. These attacks occur when an adversary has some prior knowledge about the training data and uses that knowledge to infer the sensitive data.\newline\newline
\textbf{Concern: }With a successful attack, the attacker can gain valuable information such as sensitive personal information or intellectual property.\newline\newline
\textbf{Type: }inference\newline
\textbf{Descriptor: }amplified by generative AI \newline\newline
\textbf{Implementation details: } \newline
ID: atlas-attribute-inference-attack \newline
Tag: attribute-inference-attack \newline
URI:  \href{https://www.ibm.com/docs/en/watsonx/saas?topic=SSYOK8/wsj/ai-risk-atlas/attribute-inference-attack.html}{IBM AI Risk Atlas - Attribute inference attack}\newline
\end{definitionbox}
\begin{definitionbox}{Poor model accuracy}
Poor model accuracy occurs when a model's performance is insufficient to the task it was designed for. Low accuracy might occur if the model is not correctly engineered, or there are changes to the model's expected inputs.\newline\newline
\textbf{Concern: }Inadequate model performance can adversely affect end users and downstream systems that are relying on correct output. In cases where model output is consequential, this might result in societal, reputational, or financial harm.\newline\newline
\textbf{Type: }inference\newline
\textbf{Descriptor: }amplified by generative AI \newline\newline
\textbf{Implementation details: } \newline
ID: atlas-poor-model-accuracy \newline
Tag: poor-model-accuracy \newline
URI:  \href{https://www.ibm.com/docs/en/watsonx/saas?topic=SSYOK8/wsj/ai-risk-atlas/poor-model-accuracy.html}{IBM AI Risk Atlas - Poor model accuracy}\newline
\end{definitionbox}
\begin{definitionbox}{Data transfer restrictions}
Laws and other restrictions can limit or prohibit transferring data.\newline\newline
\textbf{Concern: }Data transfer restrictions can also impact the availability of the data that is required for training an AI model and can lead to poorly represented data.\newline\newline
\textbf{Type: }training-data\newline
\textbf{Descriptor: }traditional risk of AI \newline\newline
\textbf{Implementation details: } \newline
ID: atlas-data-transfer \newline
Tag: data-transfer \newline
URI:  \href{https://www.ibm.com/docs/en/watsonx/saas?topic=SSYOK8/wsj/ai-risk-atlas/data-transfer.html}{IBM AI Risk Atlas - Data transfer restrictions}\newline
\end{definitionbox}
\begin{definitionbox}{Generated content ownership and IP}
Legal uncertainty about the ownership and intellectual property rights of AI-generated content.\newline\newline
\textbf{Concern: }Laws and regulations that relate to the ownership of AI-generated content are largely unsettled and can vary from country to country. Not being able to identify the owner of an AI-generated content might negatively impact AI-supported creative tasks.\newline\newline
\textbf{Type: }non-technical\newline
\textbf{Descriptor: }specific to generative AI \newline\newline
\textbf{Implementation details: } \newline
ID: atlas-generated-content-ownership \newline
Tag: generated-content-ownership \newline
URI:  \href{https://www.ibm.com/docs/en/watsonx/saas?topic=SSYOK8/wsj/ai-risk-atlas/generated-content-ownership.html}{IBM AI Risk Atlas - Generated content ownership and IP}\newline
\end{definitionbox}
\begin{definitionbox}{Lack of AI agent transparency}
Lack of AI agent transparency is due to insufficient documentation of the AI agent design, development, evaluation process, absence of insights into the inner workings of the AI agent, and interaction with other agents/tools/resources.\newline\newline
\textbf{Concern: }Transparency is important for AI ethics and guiding appropriate use of AI agents. Insufficient documentation might make it more difficult to govern AI agent usage, evaluate risks, to modify, or reuse the agents.  Additionally, transparency regarding how the agent's risks were determined, evaluated, and mitigated play a role in identifying an agent's suitability and evaluating its trustworthiness. The lack of standardized requirements might limit disclosure as organizations protect trade secrets and try to limit others from copying their agents.\newline\newline
\textbf{Type: }agentic\newline
\textbf{Descriptor: }amplified by agentic AI \newline\newline
\textbf{Implementation details: } \newline
ID: atlas-lack-of-ai-agent-transparency \newline
Tag: lack-of-ai-agent-transparency \newline
URI:  \href{https://www.ibm.com/docs/en/watsonx/saas?topic=SSYOK8/wsj/ai-risk-atlas/lack-of-ai-agent-transparency.html}{IBM AI Risk Atlas - Lack of AI agent transparency}\newline
\end{definitionbox}
\begin{definitionbox}{Encoded interactions attack}
Prompts that use specific encoding, styles, syntactical and typographical transformations like typographical errors or irregular spacing, or complex formatting to govern the interaction, rendering the model vulnerable.\newline\newline
\textbf{Concern: }Encoded interactions attacks can be used to alter model behavior and benefit the attacker. The content it generates may cause harms for the user or others.\newline\newline
\textbf{Type: }inference\newline
\textbf{Descriptor: }specific to generative AI \newline\newline
\textbf{Implementation details: } \newline
ID: atlas-encoded-interactions-attack \newline
Tag: encoded-interactions-attack \newline
URI:  \href{https://www.ibm.com/docs/en/watsonx/saas?topic=SSYOK8/wsj/ai-risk-atlas/encoded-interactions-attack.html}{IBM AI Risk Atlas - Encoded interactions attack}\newline
\end{definitionbox}
\begin{definitionbox}{Impact on human dignity}
If human workers perceive AI agents as being better at doing the job of the human, the human can experience a decline in their self-worth and wellbeing.\newline\newline
\textbf{Concern: }Human workers perceiving AI agents as being better at doing the humans' jobs, can cause humans to feel devalued or treated as mere data points than respected individuals. This can negatively impact society and human welfare. Reskilling can be challenging given the pace of the technology evolution.\newline\newline
\textbf{Type: }agentic\newline
\textbf{Descriptor: }amplified by agentic AI \newline\newline
\textbf{Implementation details: } \newline
ID: atlas-impact-human-dignity \newline
Tag: impact-human-dignity \newline
URI:  \href{https://www.ibm.com/docs/en/watsonx/saas?topic=SSYOK8/wsj/ai-risk-atlas/impact-human-dignity.html}{IBM AI Risk Atlas - Impact on human dignity}\newline
\end{definitionbox}
\begin{definitionbox}{Output bias}
Generated content might unfairly represent certain groups or individuals.\newline\newline
\textbf{Concern: }Bias can harm users of the AI models and magnify existing discriminatory behaviors.\newline\newline
\textbf{Type: }output\newline
\textbf{Descriptor: }specific to generative AI \newline\newline
\textbf{Implementation details: } \newline
ID: atlas-output-bias \newline
Tag: output-bias \newline
URI:  \href{https://www.ibm.com/docs/en/watsonx/saas?topic=SSYOK8/wsj/ai-risk-atlas/output-bias.html}{IBM AI Risk Atlas - Output bias}\newline
\end{definitionbox}
\begin{definitionbox}{Dangerous use}
Generative AI models might be used with the sole intention of harming people.\newline\newline
\textbf{Concern: }Large language models are often trained on vast amounts of publicly-available information that may include information on harming others. A model that has this potential must be carefully evaluated for such content and properly governed.\newline\newline
\textbf{Type: }output\newline
\textbf{Descriptor: }specific to generative AI \newline\newline
\textbf{Implementation details: } \newline
ID: atlas-dangerous-use \newline
Tag: dangerous-use \newline
URI:  \href{https://www.ibm.com/docs/en/watsonx/saas?topic=SSYOK8/wsj/ai-risk-atlas/dangerous-use.html}{IBM AI Risk Atlas - Dangerous use}\newline
\end{definitionbox}
\begin{definitionbox}{Unexplainable output}
Explanations for model output decisions might be difficult, imprecise, or not possible to obtain.\newline\newline
\textbf{Concern: }Foundation models are based on complex deep learning architectures, making explanations for their outputs difficult. Inaccessible training data could limit the types of explanations a model can provide. Without clear explanations for model output, it is difficult for users, model validators, and auditors to understand and trust the model. Wrong explanations might lead to over-trust.\newline\newline
\textbf{Type: }output\newline
\textbf{Descriptor: }amplified by generative AI \newline\newline
\textbf{Implementation details: } \newline
ID: atlas-unexplainable-output \newline
Tag: unexplainable-output \newline
URI:  \href{https://www.ibm.com/docs/en/watsonx/saas?topic=SSYOK8/wsj/ai-risk-atlas/unexplainable-output.html}{IBM AI Risk Atlas - Unexplainable output}\newline
\end{definitionbox}
\begin{definitionbox}{Human exploitation}
When workers who train AI models such as ghost workers are not provided with adequate working conditions, fair compensation, and good health care benefits that also include mental health.\newline\newline
\textbf{Concern: }Foundation models still depend on human labor to source, manage, and program the data that is used to train the model. Human exploitation for these activities might negatively impact the society and human welfare. \newline\newline
\textbf{Type: }non-technical\newline
\textbf{Descriptor: }amplified by generative AI \newline\newline
\textbf{Implementation details: } \newline
ID: atlas-human-exploitation \newline
Tag: human-exploitation \newline
URI:  \href{https://www.ibm.com/docs/en/watsonx/saas?topic=SSYOK8/wsj/ai-risk-atlas/human-exploitation.html}{IBM AI Risk Atlas - Human exploitation}\newline
\end{definitionbox}
\begin{definitionbox}{Toxic output}
Toxic output occurs when the model produces hateful, abusive, and profane (HAP) or obscene content. This also includes behaviors like bullying.\newline\newline
\textbf{Concern: }Hateful, abusive, and profane (HAP) or obscene content can adversely impact and harm people interacting with the model.\newline\newline
\textbf{Type: }output\newline
\textbf{Descriptor: }specific to generative AI \newline\newline
\textbf{Implementation details: } \newline
ID: atlas-toxic-output \newline
Tag: toxic-output \newline
URI:  \href{https://www.ibm.com/docs/en/watsonx/saas?topic=SSYOK8/wsj/ai-risk-atlas/toxic-output.html}{IBM AI Risk Atlas - Toxic output}\newline
\end{definitionbox}
\begin{definitionbox}{Unexplainable and untraceable actions}
Explanations, lineage and trace information, and source attribution for AI agent actions might be difficult, imprecise or unobtainable. \newline\newline
\textbf{Concern: }Without clear explanations, lineage trace information, and source attributions for AI agent actions, it is difficult for users, model validators, and auditors to understand and trust the model. Wrong explanations might lead to over-trust.\newline\newline
\textbf{Type: }agentic\newline
\textbf{Descriptor: }amplified by agentic AI \newline\newline
\textbf{Implementation details: } \newline
ID: atlas-unexplainable-untraceable-actions \newline
Tag: unexplainable-untraceable-actions \newline
URI:  \href{https://www.ibm.com/docs/en/watsonx/saas?topic=SSYOK8/wsj/ai-risk-atlas/unexplainable-untraceable-actions.html}{IBM AI Risk Atlas - Unexplainable and untraceable actions}\newline
\end{definitionbox}
\begin{definitionbox}{Data poisoning}
A type of adversarial attack where an adversary or malicious insider injects intentionally corrupted, false, misleading, or incorrect samples into the training or fine-tuning datasets.\newline\newline
\textbf{Concern: }Poisoning data can make the model sensitive to a malicious data pattern and produce the adversary's desired output. It can create a security risk where adversaries can force model behavior for their own benefit.\newline\newline
\textbf{Type: }training-data\newline
\textbf{Descriptor: }traditional risk of AI \newline\newline
\textbf{Implementation details: } \newline
ID: atlas-data-poisoning \newline
Tag: data-poisoning \newline
URI:  \href{https://www.ibm.com/docs/en/watsonx/saas?topic=SSYOK8/wsj/ai-risk-atlas/data-poisoning.html}{IBM AI Risk Atlas - Data poisoning}\newline
\end{definitionbox}
\begin{definitionbox}{Unreliable source attribution}
Source attribution is the AI system's ability to describe from what training data it generated a portion or all its output. Since current techniques are based on approximations, these attributions might be incorrect.\newline\newline
\textbf{Concern: }Low-quality attributions make it difficult for users, model validators, and auditors to understand and trust the model.\newline\newline
\textbf{Type: }output\newline
\textbf{Descriptor: }specific to generative AI \newline\newline
\textbf{Implementation details: } \newline
ID: atlas-unreliable-source-attribution \newline
Tag: unreliable-source-attribution \newline
URI:  \href{https://www.ibm.com/docs/en/watsonx/saas?topic=SSYOK8/wsj/ai-risk-atlas/unreliable-source-attribution.html}{IBM AI Risk Atlas - Unreliable source attribution}\newline
\end{definitionbox}

\end{document}